\documentclass[article,nofootinbib]{revtex4-1}
\usepackage[applemac]{inputenc}
\usepackage{amsmath}
\usepackage{amsfonts}
\usepackage{amssymb}
\usepackage{hyperref}
\usepackage{mathtools}
\usepackage{tikz}
\usepackage{graphicx}
\usepackage{subfigure}
\usepackage{footnote}

\newcommand{\be}{\begin{equation}}
\newcommand{\ee}{\end{equation}}

\begin{document}

\title{Neutrino flavor oscillations in a rotating spacetime}

\author{Himanshu Swami}
\email{himanshuswami@iisermohali.ac.in}
\email{ph16084@iisermohali.ac.in}
 \affiliation{Department of Physical Sciences, Indian Institute of Science Education \& Research (IISER) Mohali, Sector 81 SAS Nagar, Manauli PO 140306 Punjab India.}

\begin{abstract}
We study neutrino oscillations in a rotating spacetime under the weak gravity limit for the trajectories of neutrinos which are constrained in the equatorial plane. Using the asymptotic form of the Kerr metric, we show that the rotation of the gravitational source non-trivially modifies the neutrino phase. We find that the oscillation probabilities deviate significantly from the corresponding results in the Schwarzschild spacetime when neutrinos are produced near the black hole (still in the weak-gravity limit) with non-zero angular momentum and detected on the same side, i.e., the non-lensed neutrino.  Moreover, for a given gravitational body and  geometric parameters, there exists a distance scale for every energy scale (and vice versa), after which the rotational contribution in the neutrino phase becomes significant.
 Using the sun-sized gravitational body in the numerical analysis of the one-sided neutrino propagation, we show that even a small rotation of the gravitational object can significantly change the survival or appearance events of a neutrino flavor registered by the detector, which is located on the earth. These effects are expected to be prominent for cosmological/astrophysical scenarios where neutrinos travel past by many (rotating) gravitational bodies and for large distances. Thus rotational effects of all such bodies must be incorporated in analyzing oscillations data.
\end{abstract}

\maketitle

\section{Introduction}\label{Introduction}
It is now well established that neutrinos have non-zero mass, and the neutrino flavor eigenstates are not the same as their mass eigenstates~\cite{capozzi2014status, de2018status, esteban2019global, esteban2020fate}. Due to this, a neutrino can oscillate between their flavors, a phenomenon  known as the neutrino flavour oscillation~\cite{capozzi2014status, de2018status, esteban2019global, esteban2020fate}.  Neutrino oscillations have been studied in the literature in various gravitational physical setups~\cite{dvornikov2019neutrino, stuttard2020neutrino, mavromatos2008quantum, alfaro2002quantum,  marletto2018quantum, katori2006global, diaz2009perturbative, antonelli2018neutrino, 
motie2012high, lambiase2001neutrino, sprenger2011neutrino,chang1999possible,  sorge2007neutrino,lambiase2005neutrino,mastrototaro2021neutrino, lambiase2005pulsar,
ahluwalia1996gravitationally,ahluwalia1996interpretation, grossman1997flavor, bhattacharya1999gravitationally, luongo2011neutrino, geralico2012neutrino,koutsoumbas2020neutrino, chakrabarty2022effects, blasone2020neutrino,blasone2020beta, ahluwalia2015neutrino,lambiase2020effects,lambiase2020grbs,lambiase2021neutrino,buoninfante2020neutrino, dixit2019quantum, mukhopadhyay2007gravity,khalifeh2021distinguishing, khalifeh2021using, swami2020signature, swami2021aspects, khalifeh2021distinguishing, khalifeh2021using}. For example, neutrino flavor oscillations in stochastic gravitational waves have been studied in~\cite{dvornikov2019neutrino}, while neutrino decoherence from quantum gravitational stochastic perturbations has been studied in~\cite{stuttard2020neutrino}.
Neutrino oscillation is used to probe quantum decoherence and to constrain the quantum-gravity-induced decoherence models in~\cite{mavromatos2008quantum}. In~\cite{alfaro2002quantum}, effective dynamics of spin $1/2$ particle within the framework of loop quantum gravity is obtained, and its consequence on neutrino oscillation is discussed. Quantum gravity effects and the possible violation of Lorentz invariance  
are studied in~\cite{ marletto2018quantum, motie2012high, lambiase2001neutrino, sprenger2011neutrino,chang1999possible,katori2006global, diaz2009perturbative, antonelli2018neutrino}. 
Effects like spin-flip or helicity transitions are also studied in a gravitational setup~\cite{sorge2007neutrino,lambiase2005neutrino,mastrototaro2021neutrino, lambiase2005pulsar}.
Further, the gravitational effects on neutrino oscillations have been investigated in various backgrounds~\cite{ahluwalia1996gravitationally, ahluwalia1996interpretation, grossman1997flavor, bhattacharya1999gravitationally, luongo2011neutrino, geralico2012neutrino,koutsoumbas2020neutrino, chakrabarty2022effects}. Neutrino oscillations and effects on the beta reactions have been analyzed in the accelerated frames and in the contexts of modified theories of gravity and quintessence~\cite{blasone2020neutrino, blasone2020beta, ahluwalia2015neutrino,lambiase2020effects,lambiase2020grbs,lambiase2021neutrino,buoninfante2020neutrino}. Gravitational induced neutrino-antineutrino oscillations have also  been investigated~\cite{dixit2019quantum,mukhopadhyay2007gravity}. 
Neutrino oscillations have also been studied in the cosmological context to distinguish dark energy models and measure $H_0$ in~\cite{khalifeh2021distinguishing, khalifeh2021using}, where $H_0$ is the current value of the Hubble parameter. \\
In a previous work~\cite{swami2020signature},  neutrino lensing in the Schwarzschild geometry under the weak gravity limit has been studied where it was shown that unlike in the flat spacetime neutrino, neutrino flavor transition probability in the weak gravity, neutrino lensing contains the information about the absolute neutrino masses and their hierarchy. 
 In a curved spacetime,  multiple geodesics that connect the neutrino source and the detector can contribute to the massive neutrino wavefunction. Further, these contributions from the multiple paths in the neutrino mass eigenstates wavefunction interfere to produce a neutrino flavor transition pattern that depends on the absolute neutrino masses.
In a subsequent work~\cite{swami2021aspects}, decoherence effects in neutrino lensing arising due to the introduction of the finite width of the neutrino source and detector were studied, and decoherence length in a gravitational setup was quantified. 
Further, It was illustrated that even inferring the neutrino decoherence length can itself reveal vital information about the absolute neutrino mass.
\\
In this paper we analyze the effect of bringing in the rotation of the gravitational body into the analysis.
We study the effects of the rotation of a gravitational source on
 neutrino oscillations in regions of weak gravity to see a possible deviation in the oscillation probabilities, which may significantly change previously reported results of gravitational lensing of neutrinos in the Schwarzschild spacetime. Although neutrino oscillations in the rotating spacetime have not yet been fully explored, some limited theoretical studies of neutrino oscillations have been done in~\cite{ren2010neutrino,visinelli2015neutrino, uribe2021neutrino} for the Kerr-Neumann and Kerr spacetime. In~\cite{ren2010neutrino},  the change in the neutrino phase along the neutrino trajectories is explicitly calculated and discussed for the three following cases: (1) $L = aE$, (2) $ L=0$, in both cases, the calculations are done in the equatorial plane, and (3) the change in the neutrino phase is calculated for the neutrino radial propagation along the direction $\theta=0$, where $\theta$ is the polar angle defined in a spherical coordinate system and varies between $0\leq\theta\leq\pi$. Here, $L$, $E$ and $a$ are neutrino conserved angular momentum, neutrino conserved energy, and Kerr-Neumann rotational parameter, respectively.  In~\cite{visinelli2015neutrino},  the phase change of the neutrino in the Kerr-Neumann background is calculated in the constant $\theta$ plane for the neutrino that has zero-angular momentum ($L=0$) and is propagating in the radial direction.
In~\cite{uribe2021neutrino} neutrino flavor oscillation inside a neutrino-dominated disk around a rotating black hole is studied, and its consequences are analyzed.
 We know that the Kerr spacetime is an exact solution of the Einstein fields equation for a stationary rotating black hole with an axial symmetry around the axis of rotation. 
However, the metric for all rotating compact gravitational sources is not uniquely known and does not uniquely correspond to the Kerr metric~\cite{baines2021painleve, visser2007kerr}.
 Nevertheless, asymptotically, the metric of all compact rotating gravitational sources approximately reduce to the asymptotic metric form of the Kerr black hole~\cite{baines2021painleve, visser2007kerr}. Therefore, in this paper, we study neutrino oscillations in the weak gravity regime using the asymptotic Kerr spacetime. 
 There are many novel features of the rotating spacetime which are not present in the spherical symmetric and static spacetime. For example, the  particles can have non-planar trajectories in the rotating spacetime due to the frame-dragging~\cite{Stoghianidis1987polar,kraniotis2004precise,iyer2009light}. 
These new non-planar trajectories can now contribute to neutrino's overall wavefunction through the neutrino phase.
 Moreover, non-planar trajectories can now interfere with the other trajectories in a non-trivial manner, and  likely to modify the neutrino flavor transition probabilities.
 One expects that these deviations would be more significant in the spacetime in which a gravitational object is highly distorted~\cite{chakrabarty2022effects}.
 \\
This paper explores neutrino flavor oscillations in the asymptotic Kerr spacetime in the equatorial plane.
We show that for the one-sided propagation, i.e., non-lensed neutrino, the neutrino phase gets \textit{significantly modified} by the rotation of the gravitational source when neutrinos are produced near the gravitational source (neutrino source still remains in the weak gravity limit) and emitted non-radially. 
 We  compare the results of neutrino flavor oscillations in the asymptotics Kerr spacetime to the results in Schwarzschild spacetime.
In cases of cosmological and intergalactic  neutrinos, the deviation in the neutrino phase due to the rotation of the gravitational bodies (during  trajectory of neutrino) can  significantly change the expected flavor flux of the neutrinos received on the earth. In these cases, neutrinos travel long distances to reach the earth, and at large distances, the rotational contribution in the phase, coming due to the rotation of the gravitation bodies surrounding the trajectory of neutrinos, becomes more prominent than other contributions.
Further, the choice of $\theta =\pi/2$  in this study is argued to be justifiable in the weak gravity region for neutrinos which are only propagated either towards or away from the gravitational source. 
This is because the detector and the source which are located on the same side of the gravitational source in the equatorial plane would only receive neutrinos, which were produced and propagated only in the equatorial plane~\cite{Stoghianidis1987polar}. Hence, one does not need to worry about the non-planar trajectories of neutrinos.
Furthermore, unlike in the general neutrino propagation, neutrinos produced in the equatorial plane with initial momentum in the $\theta=\pi/2$ plane do not change their plane throughout its journey. Hence, the neutrino flux received at the detector would not be diluted as the trajectories of neutrino do not leave the plane.
However, in neutrino lensing,  the choice of the equatorial plane is not quite justifiable, where non-planar trajectories can reach  the detector that lies in the equatorial plane~\cite{Stoghianidis1987polar}. Now, the interference of non-planar trajectories with planar trajectories can non-trivially change the transition probabilities of neutrino flavors. Nevertheless, for the $\theta = \pi/2$ neutrino propagation, we show that the  rotation of the gravitational source does not significantly change the previously reported neutrino lensing results in the Schwazschild spacetime~\cite{swami2020signature}, where the local energy scale of neutrino was taken of the order $E_{loc}\sim$ MeV. Therefore, we argue that the deviation in the neutrino lensing phase due to the non-planar trajectories will not significantly  affect the results of the neutrino mass hierarchy, as previously reported in the neutrino lensing~\cite{swami2020signature, swami2021aspects}. However, there exists energy scale of neutrinos $E_{loc}\sim$ sub-KeV below which one may find a significant deviation in the neutrino lensing results for the Sun-Earth system~\cite{swami2020signature}. Further, for the supermassive black holes and galaxy clusters, one  may also find significant deviation in the lensing phase of neutrino due to the rotation of the gravitational object even for the higher energy scale is of order $E_{loc}\sim$ MeV.
 \\
This paper is organized as follows. In section \ref{sec:2},  the formalism for neutrino oscillations in the curved spacetime is briefly reviewed. In section \ref{sec:3},   we discuss the asymptotic form of the  Kerr metric in the Boyer-Lindquist coordinates. Wherein in the subsection \ref{sec:3a}, neutrino radial emission by the stationary neutrino source (local zero angular momentum observer (ZAMO) neutrino source) is discussed, followed by the subsection \ref{sec:3b}, where non-radial neutrino emission by the stationary neutrino source is discussed. In section \ref{sec:4}, numerical results are presented for the two flavor neutrino case, which is followed by the subsection \ref{decohsubsec} on  the neutrino flavor decoherence effect due to the finite neutrino and detector wavepacket. In section \ref{sec:5}, we discuss weak gravity neutrino lensing the Kerr spacetime. Finally, in section \ref{sec:6}, we conclude the results.

\section{Neutrino oscillations in curved spacetime}\label{sec:2}
Neutrino flavor oscillation formalism in the curved spacetime has been discussed for the plane wave approximation and under the weak gravity regime in \cite{cardall1997neutrino, swami2020signature}.
  The  treatment of the neutrinos as a wavepacket and its effects has been discussed in~\cite{Giunti:1991ca, giunti1998neutrinos, giunti1998coherence, grimus1999neutrino, akhmedov2009paradoxes,giunti2004coherence}, whereas  the recent treatment of the neutrino as a wave packet  has been discussed in \cite{chatelain2020neutrino, swami2021aspects, luciano2021gravitational,sadeghi2021wave}, for different curved backgrounds.   Here, we briefly overview the necessary neutrino oscillations formalism using the plane wave approximation, which is required for our present study. Neutrinos are produced in weak processes in a flavor eigenstate ($\alpha$) which is related to mass eigenstates ($i$) as
\be
|\nu_\alpha\rangle\ = \sum_{i} U_{\alpha i}^*  |\nu_i \rangle,\
\ee
where $U$ is neutrino flavor mixing unitary matrix,   $\alpha= e,\mu,\tau$ and $i=1,2,3$.
As, in a general curved spacetime, a neutrino wavefunction can have contribution from multiple geodesics which connect the neutrino source and  the detector; therefore, a neutrino flavor state propagation from source $S$ to detector $D$,  located at ${\bold x}_S$ and ${\bold x}_D$ respectively, considering all the paths, is given as \cite{swami2021aspects}
	\begin{equation}
\left|\nu_{\alpha}\left(t_{D}, \mathbf{x}_{D}\right)\right\rangle=  N_p\sum_{i} U_{\alpha i}^{*} \sum_{m}  \psi_i^{m}\left(x_D,x_S\right) \left|\nu_{i}\left(t_{S}, \mathbf{x}_{S}\right)\right\rangle,
\end{equation}
where $m$ is  the index for neutrino trajectory, which connects the neutrino source  and detector,   $\psi_i^m\left(x_D,x_S\right)$ is the wavefunction for the $i^{th}$ massive neutrino which is evolved between the time of production $(t_S)$ and detection $(t_D)$, and  $N_p$ is the overall normalization factor which in general depends on all the classically allowed trajectories ($p$) of the neutrino that connect the neutrino source and detector.  Under the plane wave approximation,  i.e.,
\be \psi_i^m\left(x_D,x_S\right)= e^{-\iota \Phi_i^m\left(x_D,x_S\right)},\ee
where $\Phi_i = \int_S^D\, p_\mu^{(i)}\, dx^\mu\,$ is the covariant phase, and  $p_\mu^{(i)}$ is the canonical conjugate momentum to the coordinate $x^\mu$ for the $i^{\rm th}$ neutrino mass, the neutrino flavor transition probability from initial produced  $\alpha$ flavor to  $\beta$  flavor at the detection point is obtained as \cite{swami2020signature}
\be \label{P_lensing}
{\cal P}_{\alpha \beta} = |\langle \nu_\beta| \nu_{\alpha}(t_D,{\bf x}_D)\rangle|^2 = |N_p|^2 
\sum_{i,j} U_{\beta i} U^*_{\beta j} U_{\alpha j} U^*_{\alpha i}\, \sum_{m,n}  \exp\left(-i \Delta \Phi^{mn}_{ij}\right)\,,
\ee
where the expression for  $|N_p|^2$ is given as~\cite{swami2020signature} 
\be \label{N}
|N_p|^2 = \left( \sum_i |U_{\alpha i}|^2 \sum_{m,n}\exp\left(-i \Delta \Phi^{mn}_{ii}\right)\right)^{-1}\,,
\ee
 with $\Delta \Phi^{mn}_{ij}= \Phi^{m}_{i}-\Phi^{n}_{j}\,$, and $m$ and $n$ are the path indices.  From Eq. \ref{N}, we see that the normalization $N_p$ depends on all the classical allowed paths of the neutrino from  the neutrino source and the detector as  the paths ($m$ and $n$ indices) are summed over.  
 
For a particular case, i.e., in a scenario when there is only one classical available path $m=1$ for the neutrino, the flavor transition probability from $\nu_\alpha \to \nu_\beta$ at the detection point becomes \cite{fornengo1997gravitational,cardall1997neutrino,swami2020signature,swami2021aspects} 
\begin{equation} \label{Prob}
{\cal P}_{\alpha \beta} \equiv \left|\langle \nu_\beta| \nu_\alpha (t_D,{\bf x}_D)\rangle \right|^2 = \sum_{i,j}U_{\beta i} U^*_{\beta j} U_{\alpha j} U^*_{\alpha i}\, \exp(-i(\Phi_i - \Phi_j))\,,
\end{equation}
where the normalisation is $N=1$. Further, for the two flavor case, the neutrino flavor transition probability from the initially produced flavor $e$ and  to the flavor $\mu$ is given as
\be \label{Twoflavour}
{\cal P}_{e\mu} = \sin^2 2\alpha\, \sin^2\left( \frac{\Delta\Phi_{12}}{2}\right),
\ee
where $\Delta \Phi_{12}= \Phi_{1}-\Phi_{2}$ and  the lepton  mixing matrix is parametrised by the mixing angle $\alpha$  as
\begin{eqnarray}
	U = \begin{pmatrix}
		\cos\alpha & \sin\alpha \\
		-\sin\alpha & \cos\alpha
	\end{pmatrix}.
\end{eqnarray}


\section{Neutrino oscillations in rotating spacetime  } \label{sec:3}
We would like to obtain the change in the neutrino phase in a background of a rotating compact star. However, the metric of a rotating compact stars is not uniquely known as there is no unique metric that corresponds to all rotating spacetime. Nevertheless, asymptotically, the metric of all compact rotating gravitational sources approximately reduces to the asymptotic metric form of the Kerr black hole~\cite{baines2021painleve, visser2007kerr}. Therefore, one can study neutrino oscillations in the weak gravity regime using the metric of an asymptotic Kerr spacetime. Hence, we will use the asymptotic form of the Kerr metric to study neutrino oscillations in the weak gravity region. The metric for the Kerr geometry in the Boyer-Lindquist coordinates  is written as
\begin{align}\label{Kerr}
ds^2 =& \left(1-\frac{2GMr}{\rho^2}\right)dt^2 +\frac{4GMar\sin^2\theta}{\rho^2}\,dt\,d\phi -\frac{\rho^2}{\Delta}\,dr^2- \rho^2 d\theta^2  -\frac{\Lambda\sin^2\theta}{\rho^2}\, d\phi^2,
\end{align}
where  $ \Delta(r) =  r^2-2GMr+a^2 $, $ \rho^2(r,\theta)= r^2+a^2\cos^2\theta$, $\Lambda(r,\theta)= \left(r^2+a^2\right)^2-a^2\Delta\sin^2\theta$ and $a=J/M$. $G$ is the Newtonian constant, $M$  is the mass and  $J$ is the angular momentum of the black hole, respectively. 
Further, the naked singularity constrains an upper bound on the $a$'s values, i.e., $a<GM$.  
Furthermore, considering neutrino propagation only in the $\theta=\pi/2$ plane ($p^\theta$=0), the metric, in the limit $a/r<<1$, ignoring $\mathcal{O}$ $(a/r)^2$, has the following  form 
\be \label{rotmetric}
ds^2\simeq B(r)\, dt^2 + 2h(r) \, dt \,d\phi-\frac{1}{B(r)}\, dr^2 -r^2\, d\phi^2\,,
\ee
where $B(r) = 1-2GM/r$ and $h(r)=2GJ/r$.  
Now, we shall discuss relativistic neutrino propagation, which is constrained in the $\theta = \pi/2$ plane. Keeping with the standard null trajectory approximation \cite{cardall1997neutrino} in literature, the integral equation for neutrino phase change from the source $r_S$ to the detector location $r_D$ for the $i^{th}$ massive neutrino, which is  produced with conserved energy $(E_0$) and angular momentum ($L$), is given by, see Appendix \ref{neutrinophase cal},
\begin{equation}\label{Eqpaper: intphaseL}
\Phi_i = \frac{m_i^2}{2}\int_{r_S}^{r_D} \frac{h(r)^2+r^2B(r)}{\left(h(r)E_0+LB(r)\right)\sqrt{ \left(B(r)\frac{E_0 r^2 -h(r)L}{h(r)E_0+LB(r) }\right)^2+ 2h(r) B(r)\frac{E_0 r^2 -h(r)L}{h(r)E_0+LB(r)} -r^2 B(r)}}dr,
\end{equation}
where $m_i$ is mass of $i^{th}$ neutrino.
Although, in general,  taking the null trajectory approximation in the neutrino phase, in the leading order of neutrino mass $m_i$, gives only a half of the neutrino phase contribution as compared to the massive neutrino trajectory, i.e., $2\Phi_i^{null} = \Phi_i^{massive}$,  see  Appendix \ref{app:nullvsmassivetrajectory}, this study will not change qualitatively if the neutrino trajectories are taken as massive trajectories.
Further, in the curved spacetime, the energy and momentum of a particle are frame-dependent, and different observers perceive different four-momenta of the particle. In a locally specified observer's frame, the local four momenta $p^{\Bar{a}}$  of the particle is related to the coordinate four momenta $p^\mu$ through local frame fields $e^{\Bar{a}}_\mu$, known as tetrad fields, by the following relation $p^{\Bar{a}} = e^{\Bar{a}}_\mu p^\mu$. 
Now, we shall first discuss a case when neutrinos are emitted radially by the stationary neutrino source. Then we will discuss when neutrinos are emitted non-radially.
 \subsection{ Radial neutrino emission  by the  stationary neutrino source }\label{sec:3a}\label{zeroL}
 Unlike in the Schwarzchild spacetime, in general, a particle cannot propagate only along the radial direction in Kerr spacetime, even if produced with zero angular momentum $L=0$, with an exception for particles traveling along $\theta=0$ line, where the metric effectively reduces to the Schwarzschild metric in the weak gravity limit.  For instance, in the asymptotic Kerr spacetime, null trajectory in $\theta=\pi/2$ plane gives 
\be\label{dphidr}
\frac{d \phi}{d r} = \frac{2GJ}{\sqrt{\left(1-\frac{2GM}{r}\right)\left(r^6-2GMr^5+4G^2r^2J^2\right)}}\, ,
\ee 
which is non-zero even for the $L=0$ case. The trajectory gets dragged along the $\phi$ coordinate due to the rotation of the gravitational source. Note the particle gets dragged in the same direction of the rotation of the gravitational source.
Further, neutrinos created with zero angular momentum as seen by the asymptotic observer will have, in general, non-zero angular momentum as seen by the local observer.
This can be seen from the relation between $p^{\phi}$ and conserved quantity $E_0$ and $L$, which is given as
 \be\label{conservedphi}
p^{\phi} = \frac{E_0h(r)+B(r)L}{h(r)^2+r^2B(r)}\Bigg|_{L=0} = \frac{E_0h(r)}{h(r)^2+r^2B(r)}\,.
\ee
As can be seen for the $L=0$, $p^\phi$ is not zero and depends on neutrino source location, except for the asymptotic observer.
However, we can make the connection between the local energy and momenta ($p^{\Bar{a}}$) of the produced neutrino with the conserved energy and angular momentum of the neutrino using local tetrad fields $e_{\mu}^{\Bar{a}}$.
Further, if neutrinos are emitted radially  in the rest frame of the neutrino source, then  only $p^{\Bar{a}}$ ($a=1,2,3$) which is along the radial direction will be non-zero. For example, in the $L=0$ case, the local frame source which satisfies such conditions, in the Kerr spacetime, is known as zero angular momentum observer (ZAMO) or stationary neutrino source. These are the special class of observers whose angular momentum ($L_{O}$) vanishes. Also, the advantage of using the local frame of the neutrino source to quantify the local energy of neutrinos is that it keeps the information about the energy scale of the neutrino processes intact at different locations in spacetime.
For example, let us say, in a neutrino production process, neutrinos are  produced with energy $E_{loc}$ by a neutrino source, which is located at $r_S$. Now, suppose the same process of neutrino production is repeated when the source is located at $r_S^{'}$. 
Then the local energy of the produced neutrinos would still be $E_{loc}$. However, for example, the neutrino energy as measured by the asymptotic observer $E_{0}$ in both the cases would be different, even though the neutrino production is the same. For example, in the Schwarzschild case, the neutrino energy as seen by the asymptotic observer would be related by $E^{'}_{0}/(1-R_s/r_S^{'})=E_0/(1-R_s/r_S)$. 
Therefore, it is more informative to write neutrino energy and momenta in terms of the local variables of the neutrino source frame. 
 Explicitly, the ZAMO tetrads~\cite{kunst2016comparing, barausse2009hamiltonian}, in the weak gravity limit can be  found as
\begin{eqnarray}
\begin{pmatrix}
e_{\mu}^{\Bar{0}} \\
e_{\mu}^{\Bar{1}} \\
e_{\mu}^{\Bar{2}} \\
e_{\mu}^{\Bar{3}}
\end{pmatrix}
= \begin{pmatrix}
   \rho\sqrt{\frac{\Delta }{\Lambda}} & 0 & 0 & 0  \\
0 & \frac{\rho}{ \sqrt{\Delta}} & 0 & 0 \\
0 & 0 & \rho & 0\\
- \frac{2 J r \sin \theta}{\rho\sqrt{\Lambda }} & 0 & 0 &\frac{\sin \theta}{\rho} \sqrt{\Lambda} \\
\end{pmatrix}
\simeq \begin{pmatrix}
   \sqrt{B(r)} & 0 & 0 & 0  \\
0 & \frac{1}{\sqrt{B(r)}} & 0 & 0 \\
0 & 0 & r & 0\\
- \frac{2 J G   \sin \theta}{r^2} & 0 & 0 &r\sin \theta \\
\end{pmatrix}\, ,
\end{eqnarray}
where,  $\mu$ can take values from $t,r,\theta,\phi$. These tetrad fields satisfies the following condition   
\begin{equation}\label{tetradconditions}
    g_{\mu\nu}=\eta_{\Bar{a}\Bar{b}}e_{\mu}^{\Bar{a}}e_{\nu}^{\Bar{b}},
\end{equation}
where $\eta_{\Bar{a}\Bar{b}}$ is dig($1,-1,-1,-1$) $4\times4$ matrix and $g_{\mu\nu}$ is the metric tensor.
 We note that the asymptotic observer also falls under the class of ZAMO observers. Further,  using $p^t$ and $p^\phi$ in terms of conserved quantities $E_0$ and $L$, i.e.,
\begin{align}\label{bothconserved}
p^{t} = \frac{E_0 r^2 -h(r) L}{h(r)^2+r^2B(r)}\,,\nonumber \\
p^{\phi} = \frac{E_0h(r)+B(r)L}{h(r)^2+r^2B(r)}\, .
\end{align}
We get local 4-momentum for the neutrino which is constrained in $\theta=\pi/2$ plane and created with $L=0$ as
\begin{align}\label{locrelationL0}
    E_{loc}(r_S) \equiv p^{\Bar{0}}= \sqrt{B(r_S)}p^t(r_S)\,, \nonumber \\ p^{\Bar{1}}=\frac{p^r(r_S)}{\sqrt{B(r_S)}}\,, \,\,\,\,
    p^{\Bar{2}}= 0\,,  \,\,\,\, p^{T}_{loc} \equiv p^{\Bar{3}}=0\,,
\end{align}
where $E_{loc}$, $p^{\Bar{1}}$, $p^{\Bar{2}}$  and $p^{T}_{loc}$ are the local energy, local momentum along the radial direction, local momentum perpendicular to $\theta=\pi/2$ plane and local tangential momentum of the neutrino respectively.
To move further, taking $L=0$ in Eq. \ref{Eqpaper: intphaseL}, we get the phase change for the $i^{th}$ neutrino, considering $r_D>r_S$,  see Appendix \ref{neutrinophase cal},
\be\label{intradialphase}
\Phi_i = \frac{m_i^2}{2E_0}\int_{r_S}^{r_D}  \frac{(\frac{2GJ}{r})^2+r^2 \left(1-\frac{2GM}{r}\right)}{\sqrt{\left(1-\frac{2GM}{r}\right)\left(r^4-2GMr^3+4G^2J^2\right)}}dr,
\ee
where, using Eq. (\ref{locrelationL0}), $E_0$ is related to $E_{loc}$ as
\begin{equation}\label{E0Eiocequation}
    E_0= E_{loc}\frac{h(r_S)^2+r_S^2B(r_S)}{r_S^2\sqrt{B(r_S)}}.
\end{equation}
Further, the neutrino phase change, in the leading order of $J$ and $M$ after ignoring higher terms such as ${\cal{O}}\left(G^4M^2J^2/r^5\right) $, is given by
\begin{align}\label{radialphase}
\Phi_i \simeq & \frac{m_i^2}{2E_0}\left[\left(r_D-r_S\right)-\frac{2G^2J^2}{3}\left(\frac{1}{r_D^3}-\frac{1}{r_S^3}\right)-G^3MJ^2\left(\frac{1}{r_D^4}-\frac{1}{r_S^4}\right) +{\cal{O}}\left(\frac{G^4M^2J^2}{r_{S,D}^5}\right)\right].
\end{align}  Here, we note that the neutrino phase depends explicitly on the mass of the gravitational source if $J\neq0$. This is in contrast to the Schwarzschild case,  where the neutrino phase, for the zero angular momentum case, depends only on the radial coordinate difference, and the gravitational dependence appeared only due to proper spatial distance of the neutrino trajectory~\cite{fornengo1997gravitational, swami2021aspects}. Now, the phase written in term of proper spatial distance ($L_p$) of the neutrino trajectory, under null approximation, can be written as 
 \begin{align}\label{properradialphase}
\Phi_i \simeq & \frac{m_i^2}{2E_0}\left[L_p -\frac{R_s}{2}\ln\frac{r_D}{r_S}+\frac{G^3MJ^2}{2}\left(\frac{1}{r_D^4}-\frac{1}{r_S^4}\right) +{\cal{O}}\left(\frac{G^4M^2J^2}{r_{S,D}^5}\right)\right],
\end{align} 
where $L_p$ is given 
\begin{align}\label{zeroLproperlength}
    L_p=  \int_{r_S}^{r_D}\sqrt{\frac{1}{B(r)}+\left(r\frac{d\phi}{dr}\right)^2}dr  \,  \simeq  & \, \underbrace{r_D-r_S + \frac{R_s}{2}\ln\frac{r_D}{r_S}}_{\text{$L_p^{Schwarzschild}$}} -\frac{2G^2J^2}{3}\left(\frac{1}{r_D^3}-\frac{1}{r_S^3}\right)  -  &\frac{3G^3MJ^2}{2}\left(\frac{1}{r_D^4}-\frac{1}{r_S^4}\right)  +{\cal{O}}\left(\frac{G^4M^2J^2}{r_{S,D}^5}\right),
\end{align}
 We notice that the neutrino phase differs from the neutrino phase in Schwarzschild geometry in two aspects (1) The spatial distance ($L_p$) between the neutrino source and the detector, which is connected by the neutrino trajectory, is more as compared  to the Schwarzschild spacetime due to the frame dragging along the $\phi$ coordinate, (2) The phase, Eq. (\ref{properradialphase}), itself gets modified by the Kerr rotational terms $MJ^2/r^4$, whereas the logarithmic term in Eq. (\ref{properradialphase}) remains the same as before for the Schwarzschild case. As both contributions in the phase are positive, for the $p^{T}_{loc}=0$ case, both aspects compel the neutrino phase change in the Kerr geometry to be more than in the Schwarzschild geometry. Though the neutrino phase gets non-trivially modified due to gravitational source rotation, this effect, however, is of order ${\cal{O}}((GJ)^2/r^4)$,  and hence, phenomenologically expected to be a minor correction in  weak gravity.


\subsection{ Non-radial neutrino emission by the  stationary neutrino source} \label{sec:3b} \label{nonzeroL}
\begin{figure}[!ht]
    \begin{center}
        \begin{tikzpicture}[scale=2]
            \draw[-] (-0.5,0) -- (4.5,0);
            
            \draw[thick,->] (0.5, 1.5)--(1, 5/3-1/3);
              \draw[thick,->] (0.5, 1.5)--(1, 5/3-1/3);  
            
                \draw[-] (0,-0.5) -- (0,15/8);
          
           \draw[->,black,dashed,bend left=14] (0.5,1.5) to (3.8,-0.5) ;
             \draw[-] (0,0) -- (3.8,-0.5);
           
           \draw[-] (0,0) -- (0.5,1.5);

\draw[ >->] (0.1,0) arc (0:180+360:0.1);
          
           \draw[black,fill=black] (0,0) circle (.05);
          
           \node[label=below: Kerr mass](4) at (0,0) {};
            \node[label=above: Source ](5) at (0.5,1.5) {};
            \node[label=below: Detector](6) at (3.8,-0.5) {};
              
                   \node[label=above:$y$](11) at (0,15/8) {};
                   \node[label=right:$x$](12) at (4.5,0) {};
                   
                     \node[label=right:$D$](14) at (3.8,-0.5) {};
                       
                        \node[label=below:$r_S$](15) at (0.5,1) {};
                         \node[label=below:$r_D$](15) at (2,-0.5) {};
        \end{tikzpicture}
    \end{center}
    \caption{ Diagrammatic representation of the neutrino source and the detector setup, when neutrinos are produced with a non-zero angular momentum $L\neq0$ and propagated away from the gravitational source. Here, $J$ and $L$ have different direction of angular momentum. }
    \label{fig:setup}
\end{figure}
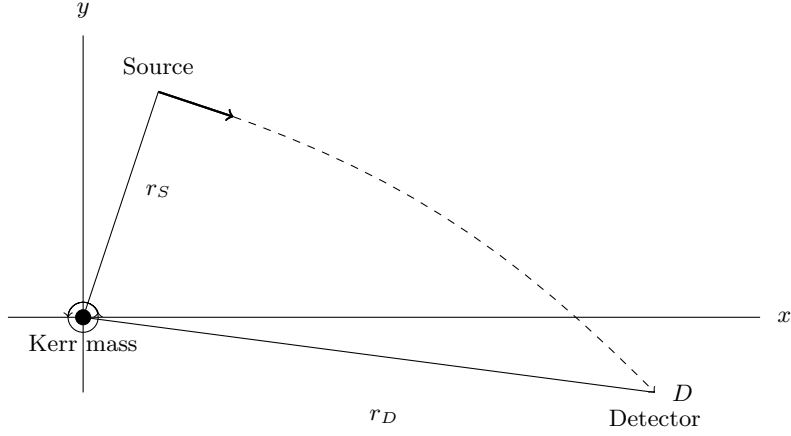
In this section, we consider neutrinos produced at $\vec{r}_S$ as a plane wave with conserved energy $E_0$  and angular momentum $L$  and detected by detector at $\vec{r}_D$, which lies on the same side as the neutrino source, see Fig. \ref{fig:setup}. 
 As previously noted, angular momentum of a particle is observer dependent and related to the local tetrad fields. In the case when neutrinos are produced with non-zero conserved angular momentum ($L\neq0$),  a source frame tetrads, in which  neutrinos are produced (in a neutrino production process) along the radial direction, can be found using tetrads fields constraints imposed by the following relations
\begin{align}
    p^{\Bar{2}} =  \widetilde{e}_{\mu}^{\Bar{2}}p^{\mu}=0 \,, \nonumber \\
   p^{\Bar{3}} =  \widetilde{e}_{\mu}^{\Bar{3}}p^{\mu}=0\,,
\end{align}
whereas $e^{\Bar{1}}_{\mu}$ is such that, if  $p^r=0$, then this should imply $  p^{\Bar{1}}=0$. This ensures the $e^{\Bar{1}}_{\mu}$ is locally directed along   the radial direction only. 
For instance, if a local neutrino source has the following tetrads fields $ \widetilde{e}_\mu^{\Bar{0}}=(e_{t}^{\Bar{0}}, 0, 0, e_{\phi}^{\Bar{0}})$, $\widetilde{e}_\mu^{\Bar{1}}=(0 , e_{r}^{\Bar{1}} , 0 , 0)$, $\widetilde{e}_\mu^{\Bar{2}}=(0 , 0 , e_{\theta}^{\Bar{2}} , 0)$, and $\widetilde{e}_\mu^{\Bar{3}}=(e_{t}^{\Bar{3}}, 0 , 0 , e_{\phi}^{\Bar{3}})$.
Then for the neutrino trajectories constrained in $\theta=\pi/2$ plane ($p^{\theta}=0$), we have the following constraint
\begin{eqnarray}
p^{\Bar{3}} = 0 = e_{\phi}^{\Bar{3}}p^{\phi}+ e_{t}^{\Bar{3}}p^t \Rightarrow \frac{e_{\phi}^{\Bar{3}}}{e_{t}^{\Bar{3}}}=-\frac{p^t}{p^{\phi}}= -\frac{E_0r^2-h(r)L}{E_0h(r)+B(r)L}\Bigg|_{r=r_S},
\end{eqnarray}
which can be written as 
\begin{equation}\label{constraintdiagonal}
    \frac{L}{E_0} = \frac{e_{t}^{\Bar{3}}r^2+e_{\phi}^{\Bar{3}}h(r)}{h(r)e_{t}^{\Bar{3}}-B(r)e_{\phi}^{\Bar{3}}}\Bigg|_{r=r_S}.
\end{equation}
The above equation relates the frame of the source with the conserved neutrino energy and angular momentum. Further, these new tetrads are related to ZAMO tetrads by a local Lorentz boost along $x^{\bar{3}}$ direction $\widetilde{e}_\mu^{\bar{a}}= \Lambda_{\bar{b}}^{\bar{a}}e_\mu^{\bar{b}}$. Lorentz matrix elements  ($\Lambda_{\bar{b}}^{\bar{a}}$)  can be found out from Eq. (\ref{constraintdiagonal}) in terms of $L$ and $E_0$.   Moreover, all observers' satisfying Eq. \ref{constraintdiagonal} at their location  will perceive neutrinos as if they are emitted only along the radial direction.
%
We, however,  take the ZAMO observer's frame (stationary neutrino source) as the local neutrino source frame. 
Now, the phase change for the massive neutrino $i$ during the trajectory from $\vec{r}_S$ to $\vec{r}_D$ ($\vec{r}_D>\vec{r}_S$), and up to  the first leading   terms in mass $M$ and the angular momentum $J$ is given by, see Appendix \ref{neutrinophase cal},
\begin{equation}\label{Phase}
\Phi_i\simeq \frac{m_i^2}{2E_0}\left[\sqrt{r_D^2-\left(\frac{L}{E_0}\right)^2} -\sqrt{r_S^2-\left(\frac{L}{E_0}\right)^2}
+  \left(GM-\frac{2GJE_0}{L}\right)\left(\frac{r_D}{\sqrt{r_D^2-\left(\frac{L}{E_0}\right)^2}}-\frac{r_S}{\sqrt{r_S^2-\left(\frac{L}{E_0}\right)^2}}\right)\right].
\end{equation} 
 The above expression of the neutrino phase can be reduced to the $L=0$ case by first expanding the neutrino phase in terms of dimensionless term $L/(E_0 r_{D,S})$ and then taking the limit $L\rightarrow 0$. Also, as the neutrino phase, Eq. \ref{Phase} is written in the leading order of $GM$ and $GJ$; the $L\rightarrow0$ limit of  Eq. \ref{Phase} will not contain terms of $\cal{O}$ $(G^2J^2)$ as given in the Eq. \ref{radialphase}.  
Now $E_0$ and $L/E_0$  are given in terms  of neutrino local energy $E_{loc}$ and local tangential momentum $p^{\bar{3}}\equiv p^{T}_{loc}$  as
\begin{align}\label{E0/L0localequation}
E_{0} &  =  \frac{E_{loc}\left(h(r_S)^2+r_S^2B(r_S)\right)}{\sqrt{B(r_S)}\left(r_S^2-h(r_S)\frac{L}{E_0}\right)}\,,\nonumber \\
    \frac{L}{E_0} & = \frac{r_S^3p^{T}_{loc} \sqrt{B(r_S)}}{B(r_S)E_{loc}r_S^2+h(r_S)\left(h(r_S)E_{loc}+r_S\sqrt{B(r_S)}p^{T}_{loc}\right)}\,,
\end{align}
whereas the relation $p^{\bar{1}}= p^r/\sqrt{B(r_S)}$ remains same.
 We notice that phase  Eq. (\ref{Phase}), unlike in the Schwarzschild geometry, is not invariant under $p^{T}_{loc}\rightarrow-p^{T}_{loc}$. This statement is equivalently true in terms of $L$. Consequently, the neutrino flavor transition probability will depend on whether neutrinos are produced with  angular momentum $L$ aligned or anti-aligned with the rotation of  Kerr black hole. 
Further, writing the phase in term of the proper spatial distance  of the neutrino trajectory, under the null approximation,  we get
\begin{equation}\label{properphaseLnotzero}
    \Phi_i= \frac{m_i^2}{2E_0}\left[ L_p - GM\left(\tanh^{-1}\sqrt{1-\left(\frac{L}{E_0 r_D}\right)^2}-\tanh^{-1}\sqrt{1-\left(\frac{L}{E_0 r_S}\right)^2}\right)\right],
\end{equation}
where 
\begin{equation}
\begin{aligned}\label{properdistancenonradial}
    L_p = &\int_{r_S}^{r_D}\sqrt{\frac{1}{B(r)}+\left(r\frac{d\phi}{dr}\right)^2}dr  \,  \simeq   \sqrt{r_D^2-\left(\frac{L}{E_0}\right)^2} -\sqrt{r_S^2-\left(\frac{L}{E_0}\right)^2} + GM\Bigg(\tanh^{-1}\sqrt{1-\left(\frac{L}{E_0 r_D}\right)^2}  - \\ & \tanh^{-1}\sqrt{1-\left(\frac{L}{E_0 r_S}\right)^2}   
     + \frac{r_D}{\sqrt{r_D^2-\left(\frac{L}{E_0 }\right)^2}}-\frac{r_S}{\sqrt{r_S^2-\left(\frac{L}{E_0 }\right)^2}} \Bigg) - \frac{2G JE_0}{L}\left(\frac{r_D}{\sqrt{r_D^2-\left(\frac{L}{E_0 }\right)^2}}- \frac{r_S}{\sqrt{r_S^2-\left(\frac{L}{E_0 }\right)^2}}\right),  
\end{aligned}
\end{equation}
we used the following relation to calculate $L_p$
\begin{equation}
    \frac{dr}{d\phi}= \sqrt{\left(B(r)\frac{p^t}{p^\phi}\right)^2+4B(r)\frac{GJ}{r}\frac{p^t}{p^\phi}-B(r)r^2}\, \,\,.
\end{equation}
The Eq. \ref{properdistancenonradial} can be shown to reduce to the $L\rightarrow0$ limit, i.e., Eq. \ref{zeroLproperlength} (up to $R_s$ order)   by first using Logarithmic Hyperbolic inverse identity  ($\tanh^{-1}x = 1/2\ln|(1+x)/(1-x)|$), then expanding the phase and Logarithmic argument  in terms of $L/E_0(r_{S,D})$ dimensionless parameter and then taking $L\rightarrow0$ limit.
Further, unlike the Schwarzschild case,  $L_p$ is not invariant under $p^{T}_{loc}\rightarrow-p^{T}_{loc}$ (using $L/E_0$ relation from Eq. \ref{E0/L0localequation}), i.e., now, the spatial distance travelled by the particle  depends on the  the particle tangential momentum direction.  This feature is due to frame dragging of the particle.
Furthermore, we note that neutrino phase's first explicit dependence on $J$, other than from $L_p$, appears at ${\cal{O}}$ $(G^2JME_0^2/(L^2))$, and given as, 
\begin{eqnarray}\label{Eq:tidelphasenextorder}
    \tilde{\Phi_i} = & \frac{m_i^2}{2E_0}\Bigg[ \tilde{L_p} - GM\Bigg(\tanh^{-1}\sqrt{1-\left(\frac{L}{E_0 r_D}\right)^2}-\tanh^{-1}\sqrt{1-\left(\frac{L}{E_0 r_S}\right)^2}\Bigg) - \nonumber\\ &\frac{2G^2JME_0^2}{L^2}\left(\sin^{-1}\left(\frac{L}{E_{0} r}\right)\bigg|_{r_S}^{r_D}- \frac{L}{E_0\left(r^2-\frac{L^2}{E_0^2}\right)^{1/2}}\Bigg|_{r_S}^{r_D}\right)\Bigg], 
\end{eqnarray}
where the corresponding proper spatial distance $\tilde{L_p}$  is given by
\begin{equation}\label{Eq:propertidellength}
    \tilde{L_p} = L_p-\frac{2G^2JME_0^2}{L^2}\left[\frac{L\left(4r^2-5\frac{L^2}{E_0^2}\right)}{E_0\left(r^2-\frac{L^2}{E_0^2}\right)^{3/2}}\Bigg|_{r_S}^{r_D}-4\sin^{-1}\left(\frac{L}{E_{0} r}\right)\bigg|_{r_S}^{r_D}\right] . 
\end{equation}
Again, the Eq. \ref{Eq:tidelphasenextorder} and Eq. \ref{Eq:propertidellength} can be shown to reduce to the $L\rightarrow0$ limit by first using Logarithmic Hyperbolic inverse identity  ($\tanh^{-1}x = 1/2\ln|(1+x)/(1-x)|$), then expanding the phase and Logarithmic argument  in terms of $L/E_0(r_{S,D})$ dimensionless parameter and then taking $L\rightarrow0$ limit.
Also, note that  ${\cal{O}}$ $(G^2JME_0^2/(L^2))$ is subdominant and  suppressed term in the weak gravity. Hence, effectively in the weak gravity limit, for an asymptotic observer, the distinction between the neutrino phase in the Schwarzschild and the asymptotic Kerr spacetime emerges only from the difference in the proper spatial length along the neutrino trajectory which connects the neutrino source and detector. 
The changes in the proper oscillation length of neutrino along with the other terms in the neutrino phase, which are brought up in the neutrino phase due to the rotating mass, can nevertheless be significant, which we will show in the following section using numerical examples.

\section{Numerical results}\label{sec:4}
\begin{figure}
\includegraphics[width=1\textwidth]{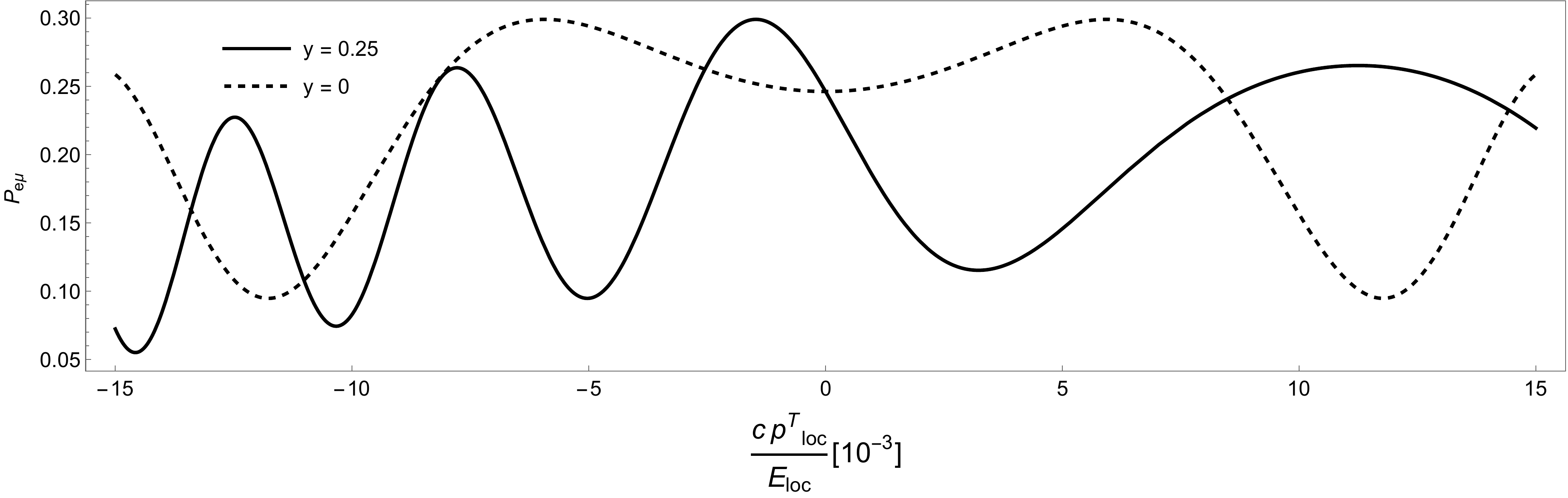}
\caption{Probability $\nu_e \to \nu_\mu$ conversion in the three flavor neutrino case as a function of $cp^{T}_{loc}/E_{loc}$ for  (1)  $y \equiv Jc/GM^2 \neq 0$, and (2) $y=0$, i.e., Schwarzschild case.  Here, $R_s = 30$ km,  $r_S = 10^5$ km, $r_D=10^
{11}$ km, and $E_{loc} = 10$ MeV.  The values of neutrino mass squared differences, mixing angle and the Dirac CP phase are taken from the  (NuFIT 5 (2020)) global fit \cite{esteban2020fate}.}  \label{fig: diff J compare}
\end{figure}
  \begin{figure}[]
\centering
\includegraphics[width=1\textwidth]{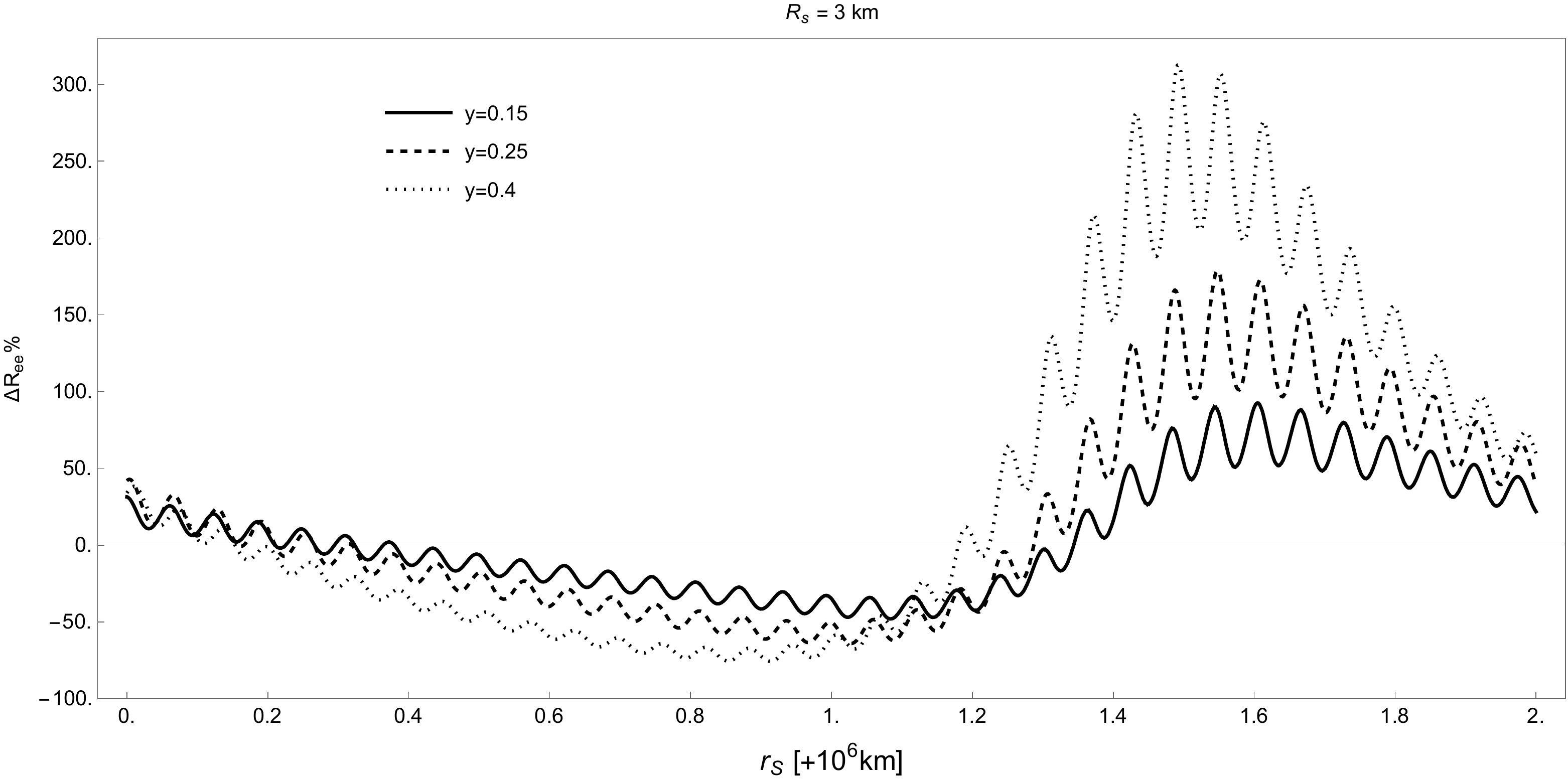}
\caption{Plot of the relative change in the survival probability of the  electron ($e$) flavor of neutrino due to the rotation of  gravitational source in the three flavor case as a function  of neutrino source location $r_S$. Here, solid black curve, black dashed curve, and black dotted curve is for the $y=0.15$,  $y=0.25$, and  $y=0.4$, spin parameters respectively, while the detector is located at $r_D = 10^{13}$ km.   Other parameters are $R_s= 3$ km, $ p^{T}_{loc}= 0.4 \, E_{loc}$ MeV/c, and $E_{loc}= 1$ MeV. The values of neutrino mass squared differences, mixing angle and the Dirac CP phase are taken from the  (NuFIT 5 (2020)) global fit \cite{esteban2020fate}. 
}
\label{fig:AlphaSolar}
\end{figure}
  \begin{figure}[]
\centering
\includegraphics[width=1\textwidth]{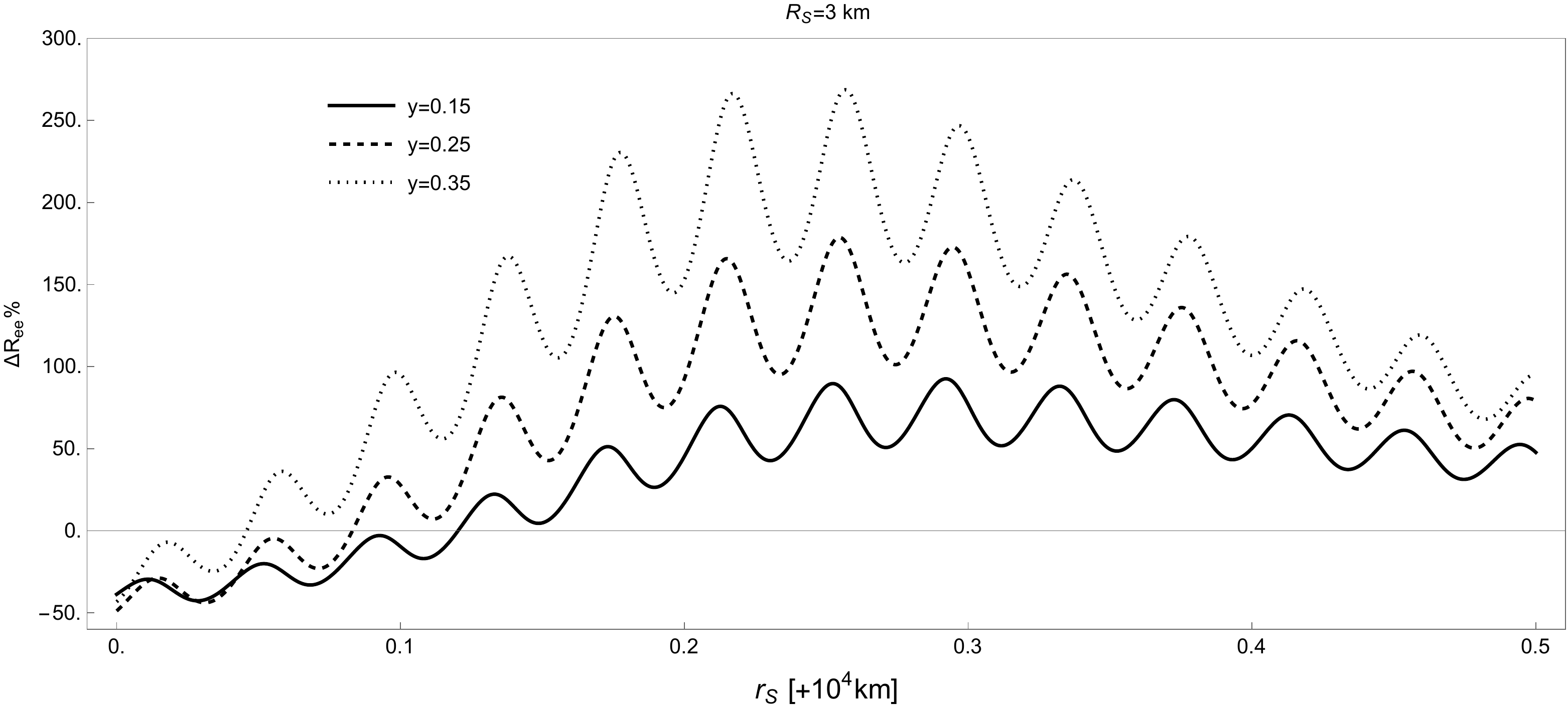}
\caption{Plot of the relative change in the survival probability of the  electron ($e$) flavor of neutrino due to the rotation of  gravitational source in the three flavor case as a function  of neutrino source location $r_S$. Here, solid black curve, black dashed curve, and black dotted curve is for the $y=0.15$,  $y=0.25$, and  $y=0.35$, spin parameters respectively,  while the detector is located at $r_D = 10^{8}$ km.  Other parameters are $R_s= 3$ km,  $ p^{T}_{loc}= 0.4 \, E_{loc}$ MeV/c, and $E_{loc}= 0.1$ MeV. The values of neutrino mass squared differences, mixing angle and the Dirac CP phase are taken from the  (NuFIT 5 (2020)) global fit \cite{esteban2020fate}. 
}
\label{fig:Suneathy0103}
\end{figure}

In this section, we demonstrate previously discussed neutrino phase modification in the asymptotic Kerr background using numerical examples and compare it with the Schwarzschild background for the three neutrino flavors case. The following probabilities expression is  used for the numerical calculation: 
\begin{equation} \label{NumProb}
{\cal P}_{\alpha \beta} \equiv \left|\langle \nu_\beta| \nu_\alpha (t_D,{\bf x}_D)\rangle \right|^2 = \sum_{i,j}U_{\beta i} U^*_{\beta j} U_{\alpha j} U^*_{\alpha i}\, \exp(-i(\Phi_i - \Phi_j))\,,
\end{equation}
with the neutrino phase difference
\begin{equation}\label{DiffNuPhase}
\Delta\Phi_{ij}= \frac{\Delta m_{ij}^2}{2E_0}\left[\sqrt{r_D^2-\left(\frac{L}{E_0}\right)^2} -\sqrt{r_S^2-\left(\frac{L}{E_0}\right)^2}
+ \left(GM-\frac{2GJE_0}{L}\right)\left(\frac{r_D}{\sqrt{r_D^2-\left(\frac{L}{E_0}\right)^2}}-\frac{r_S}{\sqrt{r_S^2-\left(\frac{L}{E_0}\right)^2}}\right)\right],
\end{equation}
where $\Delta m_{ij}^2\equiv m_{i}^2-m_{j}^2$ and $i,\,j\in {1,2,3} $, while $E_0$ and $L/E_0$ relations in terms of the local energy and tangential momentum of the neutrino are taken from Eq. \ref{E0/L0localequation}.

Further, one can  find out from  Eqs. $(27,\, 34)$, the shift in the neutrino phase for the $i^{th}$ mass eigenstate ($\Delta\Phi_{shift} \equiv |\Phi(J\neq0)-\Phi(J=0)|$)   due to the rotation of the gravitational object;  in the leading order of $R_s\equiv 2GM$ and  the  dimensionless spin parameter ($y\equiv J/GM^2$) and for  $r_S/r_D <<1$ case  is given as
\begin{equation}\label{onesideshift}
\Delta \Phi_{shift} \simeq \frac{m_i^2 \, y \, R_s^2 \, r_D  \, p^T_{loc}} {4\,  r_S^2 \, E_{loc}^2} = 1.27 \, y \left(\frac{c\,p^T_{loc}}{E_{loc}}\right)\left(\frac{R_s}{r_S}\right)^2 \left(\frac{m_i^2} {eV^2/c^4}\right) \, \frac{r_D/E_{loc}}{km/GeV},
  \end{equation}
where $c$ is the speed of light.
This can be readily used  to estimate the order of phase shift for different parameters in the astrophysical setups ($y$, $R_s$, $r_S$, $r_D$, $E_{loc}$ and $p^T_{loc}$).
For example, using the following parameters $r_D \sim 10^{13}$ km, $R_s \sim 3$ km, $m_i^2\sim 10^{-3}$ eV$^2$/c$^4$, $y=0.15$, $cp^T_{loc}/E_{loc}=0.4$, and $E_{loc}=1$ MeV ($0.001$ GeV). One gets $\Delta \Phi_{shift} \simeq 6.858$ when the neutrino source is  located at  $r_S=10^6$ km, which is $\mathcal{O}$ ($1$) or roughly a full cycle of $2\pi$. For  $r_S = 10^5$ km, the shift in the neutrino phase is  $\Delta \Phi_{shift}\simeq 685.8$, which is $\mathcal{O}$ ($10^2$) or roughly  $10^2$ cycles of $2\pi$. And for  $r_S = 10^4$ km, the shift in the neutrino phase is  $\Delta \Phi_{shift}\simeq 68580$, i.e., $\mathcal{O}$ ($10^4$) or roughly  $10^4$ cycles of $2\pi$. 
These shifts in the neutrino phase are sufficiently large to produce a significant relative difference in the transition probabilities of neutrino flavors, which we will see in the following numerical examples.
From Eq. \ref{onesideshift}, we see that as $r_D$ gets larger and larger, the phase shift also becomes larger and larger. Therefore, neutrino sources with cosmological or intergalactic origins would see a significant deviation in the flavor flux received on the earth, irrespective of the energy scales of the neutrino.
However, we will see through numerical examples of the Alpha Centauri-Earth and Sun-Earth systems that these phase shifts can also be significant even within the intragalactic distances. 

In the plots, the neutrino mixing parameters and masses are taken from the NuFIT 5 (2020)\cite{esteban2019global} for the normal ordering, 
which are $\Delta m_{21}^2 = 7.42 \times 10^{-5}~{\rm eV}^2$,  $\theta_{12}=33.44^{\circ}$, $\Delta m_{31}^2 = 2.517 \times 10^{-3}~{\rm eV}^2$, $\theta_{23}=49.2^{\circ}$, $\theta_{13}=8.57^{\circ}$, $\delta_{\rm CP}=197^{\circ}$. The $U$ matrix in three flavor case is also taken from NuFIT 5 (2020)\cite{esteban2020fate}. 
In the following plots, We use a wide range of parameters to show the importance of including the rotation of the gravitational source while analyzing the oscillation probabilities.
\\
  In the Fig. \ref{fig: diff J compare}, neutrino flavor transition probability ($\nu_e\rightarrow \nu_{\mu}$) is plotted as a function of $c p^{T}_{loc}/E_{loc}$ for different values of dimensionless spin parameter $y\equiv Jc/GM^2$, where $c$ is the speed of light.
  In the plots of Fig. \ref{fig: diff J compare}, $c p^{T}_{loc}/E_{loc}$ takes values from $-15 \times 10^{-3}$ to $15 \times 10^{-3}$.  Further, the following parameters  $E_{loc}=10$ MeV, $r_D= 10^{11}$ km, and $r_S= 10^5$ km are used in the plots.  The location of the neutrino source ($r_S$) is chosen such that the source is located in the  weak-gravity region. We take $R_s = 30$ km, which is of the $R_s\sim\mathcal{O}\left(10
\right)$. These type of gravitational sources can be massive stars as well as black holes. The shift in the neutrino phase corresponding to these parameters, using the smallest value of $\Delta m_{21}^2$, i.e., $m_i^2 \sim \Delta m_{21}^2 \sim 10^{-5} eV^2$,  is of the order $\mathcal{O}(10^2)$. \\
From the Fig. \ref{fig: diff J compare}, we notice the followings:
 \begin{itemize}
     \item The neutrino flavor transition probability is highly sensitive to the rotation of the gravitational source. For instance, the spin parameter $y=0.25$ (that corresponds to the solid curve) gives a very different neutrino flavor transition profile at the location of the detector compared to the dashed curve, which represents the Schwarzschild case. For example, at around $c p^{T}_{loc}/E_{loc}=12 \, [10^{-3}]$, the chance of appearance of the $\mu$ neutrino flavor in the Schwarzschild case ($y=0$) is around  $0.10$, whereas, for the  $y=0.25$ case, the chance of appearance of  muon ($\mu$)  flavor of neutrino is around $0.25$. Therefore, the relative change in registering events of  the appearance of the $\mu$ neutrino flavor in the $y=0.25$ case has increased to $150\%$ when compared to the Schwarzschild case ($y=0$). In contrast, around $c p^{T}_{loc}/E_{loc}=4\, [10^{-3}]$, we will see an opposite trend as the appearance probability of the muon ($\mu$) flavor of the neutrino for the $y=0.25$ case has decreased when compared to the Schwarzschild case ($y=0$).
\item Further, unlike for the Schwarzschild case ($y=0$), the probability of neutrino flavor transition is asymmetric around $c p^{T}_{loc}/E_{loc}= 0$. Consequently, the transition probability is also sensitive to the neutrino tangential momentum direction.  This  $c p^{T}_{loc}/E_{loc}\rightarrow-c p^{T}_{loc}/E_{loc}$ invariance  in the neutrino phase is broken because of the frame-dragging induced by the rotation of the spacetime.  In one case, neutrino needs to traverse along with the rotation of the spacetime, and in other case, neutrino traverse against  the rotation of the spacetime.  Consequently, we get distinct neutrino phases in these two cases.

\end{itemize}

Further,  to clearly estimate  the effects on the transition probabilities of the neutrino flavor due to the rotation of the gravitational source, we plot the relative change in the neutrino flavor transition probability ($\Delta R_{e\alpha}\% $) from the electron ($e$) flavor to the $\alpha$ type of neutrino flavor, where $\alpha$ represents any of electron $e$, muon $\mu$, or tau $\tau$  neutrino flavor.    The expression of $\Delta R_{e\alpha}(\%) $ is defined as
\be\label{ratiopercentage}
\Delta R_{e\alpha}\left(\%\right) \equiv \frac{\Delta P_{e\alpha}}{P_{e\alpha}^{Schwarzschild}}\% \equiv \frac{P_{e\alpha}(J\neq0) -P_{e\alpha}(J=0)}{P_{e\alpha}(J=0)}\times 100.
\ee
\\
We present numerical results for the Sun-sized gravitational object. In the analysis, the detector location roughly corresponds to the distance between the Sun and the Earth ($r_D=10^8$ km) and the distance between the Alpha Centauri and the Solar system ($r_D=10^{13}$ km). These calculations are general enough to be applied to more related astrophysical objects. Unlike the parameters used in the previous example, which gave an $\mathcal{O}(10^2)$ shift in the neutrino phase, in the following examples, to see the effects of rotating gravitational mass on the oscillation probabilities of neutrino, we will take parameters that only correspond to a maximum of $\mathcal{O}(1)$ shift in the neutrino phase. Using Eq. \ref{onesideshift}, the shift of an $\mathcal{O}(1)$ in the neutrino phase due to rotating Sun-sized object ($R_s \sim 3$ km), i.e., $\Delta \Phi_{shift}\sim \mathcal{O} (1)$, taking the extremal rotational value ($y=1$) and maximum value of the tangential momentum of neutrino $\left|cp^T_{loc}/E_{loc}\right| =1$,  gives the following relation 
\begin{equation}\label{order_1_phase_shift_for_sunsized}
    r_S \sim 3.38\, m_i\sqrt{\frac{r_D}{E_{loc}}},
\end{equation}
where $r_S$ and $r_D$ are in km, and $m_i$ and $E_{loc}$ are in the eV and GeV units, respectively.
A typical energy spectrum of neutrinos inside Sun-like stars ranges from a KeV scale to a MeV scale. Therefore, using Eq. \ref{order_1_phase_shift_for_sunsized}, we can get an estimation of the location of neutrino sources, after which the phase of neutrinos, which are produced with a certain energy scale $E_{loc}$ will get shifted by $\mathcal{O}\left(1\right)$ due to the rotating mass.

For the detector located around $r_D \sim 10^{13}$ km  (Alpha Centauri to Solar system distance) and using the smallest $\Delta m^2$ value, $m_i^2\sim \Delta m_{21}^2 \sim 10^{-5}$ eV$^2$, we get
\begin{eqnarray}
 r_S  \sim 10^8 \,\,\text{km} \,\,\,\,\text{for} \,\,\,\,& E_{loc}\sim \text{KeV}\, ,\nonumber\\
 r_S  \sim 10^6 \,\, \text{km} \,\,\,\, \text{for} \,\,\,\,& E_{loc}\sim \mathcal{O}(1)\,\text{MeV} \, .
\end{eqnarray}
We note the followings:
\begin{itemize}
     \item Those neutrinos produced with a KeV energy scale will have a significant shift in their phases if their production sources lie below or around $r_S \sim 10^8$ km. On the other hand, after $r_S \gtrsim 10^8$ km, the neutrino with an energy scale of $E_{loc} \sim$ KeV will not get a significant correction in the neutrino phase due to the rotating mass.
    \item For the neutrinos with an energy scale of $E_{loc}\sim$ MeV, the  production of neutrinos should occur  below the order of  $r_S  \sim 10^6$ km to get a phase shift of $\mathcal{O}(1)$.
    \item Therefore,  the highly energetic neutrinos will get less window for the locations of their productions, where their phases will get significantly shifted due to the gravitational effect of a rotating mass.
\end{itemize}
In the following plots, we  show that even an $\mathcal{O}(1)$ shift in  neutrino phases can significantly change the flavor flux of neutrinos at the detector site. 
In Fig. \ref{fig:AlphaSolar}, we have plotted the relative change in the survival probability of the $e$ neutrino flavour as  a function of the neutrino source location $r_S$ to see the impact on the survival probability of the electron ($e$) flavor of neutrino due to the rotation of the gravitational source, i.e., now we take $\alpha=e$ in the   $\Delta R_{e\alpha}(\%) $ expression, Eq. \ref{ratiopercentage}. 
 Further, we take  the Schwarzschild radius as $R_s=3$ km (which might be relevant for the Sun and Sun sized stars, e.g., the Alpha Centauri) and different spin parameters in the Fig. \ref{fig:AlphaSolar}.
 In the plots of Fig.  \ref{fig:AlphaSolar},  local energy of neutrino,  local tangential momentum of neutrino and the  location of the detector are taken as $E_{loc} = 1$ MeV, $ p^{T}_{loc}= 0.4 \, E_{loc}$ MeV/c and $r_D =10^{13}$ km (which approximately corresponds the distance between the Solar and the Alpha Centauri systems),  respectively, while $r_S$ is changed from $10^6$ km to $2 +10^6 $ km. These values correspond to the maximum of $\mathcal{O}(1)$ shift in the neutrino phases.
  The positive values of $\Delta R_{e\alpha}(\%)$, e.g., $\Delta R_{e\alpha}(\%)$ = $100\%$ in the plots represent a $100\%$ jump in the survival rate of the $e$ flavor at the detector compared to the non-rotating case, whereas negative values of  $\Delta R_{e\alpha}(\%)$ mean that the chance of the survival of  $e$  flavor has decreased at the detector in the rotating case  compared to the non-rotating case.  The solid black curve, dashed black curve, and black dotted curve in  Fig. \ref{fig:AlphaSolar} is plotted for the different spin parameters $y\equiv Jc/GM^2$ and  their values are $y=0.15$, $y=0.25$, and $y=0.4$, respectively.
\\
\\
\\
From the Fig. \ref{fig:AlphaSolar}, we make the following  observations:
\begin{itemize}
    \item In Fig. \ref{fig:AlphaSolar}, we see a significant  relative change in the survival probabilities ($\Delta R_{ee}\%$) due to the rotation of  the gravitational source  in all the cases of  $y$ ($y= 0.15,\, 0.25, \, \& \, 0.4$).   For example, for the location of the neutrino source, which lies between $r_S\, =1.4 \, [+10^6]$ km to $2 \, [+10^6]$ km,  we see that the relative change in the survival probability of the electron flavor in all the $y$ cases is positive. Hence, in all the $y$ cases, the detector is more likely to find the $e$ flavor neutrino in a rotating spacetime than by a detector which is placed in a non-rotating spacetime.  From location,  $r_S\, =0.4 \, [+10^6]$ km to $r_S\, =1.2 \, [+10^6]$ km,  we see a negative value of the relative change in the survival probability of the electron flavor in all the $y$ cases. Hence, the detector is less likely to register the survival events of  $e$ flavor in a rotating spacetime with spin parameters $y=0.15,\,0.25,\,\& \,0.4$ compared to a detector, which is placed in a non-rotating spacetime. 
\item Further, we clearly see the distinction  between the black solid curve of  $y= 0.15$, black dashed curve of $y=0.25$, and blacked dotted curve of $y=0.4$. For example, around $r_S= 1.6 \, [+10^6]$ km, the value of $\Delta R_{ee}\%$ for $y=0.4$ fluctuate around $250\%$, while for $y=0.15$, the values of $\Delta R_{ee}\%$ fluctuate around $50\%$. Therefore, the survival events registered by the detector will be quite different in each of the cases of $y$, i.e., even for different moderate rotations.  

\end{itemize}
To further illustrate the effects of the rotating mass on neutrino oscillations, we take the Sun and the Earth system to plot the relative change in the survival probability of the electron ($e$) flavor of neutrino as a function of the neutrino source location $r_S$. 
 Similarly, as done in the previous example, we estimate the  $\mathcal{O} (1)$ shift in the neutrino phase due to the rotation of Sun using Eq. \ref{order_1_phase_shift_for_sunsized}.
The energy spectrum of neutrinos emitted by the Sun ranges from a KeV scale to an $\mathcal{O}(10)$ MeV scale~\cite{bahcall200610}. 
\\
For the detector located around $r_D \sim 10^{8}$ km  (approximately the distance between the Sun and the Earth) and with $m_i^2\sim \Delta m_{21}^2 \sim 10^{-5}$ eV$^2$, we get
\begin{eqnarray}\label{Sun-Earth_table_for_energy_scale}
r_S  \sim  10^4 \,\,\text{km} \,\,\,\,\text{for} \,\,\, & E_{loc}\sim \text{KeV}\, ,\nonumber\\
 r_S  \sim  10^3 \,\, \text{km} \,\,\,\, \text{for} \,\,\,\, & E_{loc}\sim \mathcal{O}(1)\,\text{MeV} \,,\nonumber\\
 r_S  \sim  10^2 \,\, \text{km} \,\,\,\, \text{for} \,\,\,\, & E_{loc}\sim \mathcal{O}(10)\,\text{MeV} \, .
\end{eqnarray}
We note the followings:
\begin{itemize}
     \item  Neutrinos produced with the  KeV, $\mathcal{O}(1)$MeV, and $\mathcal{O}(10)$MeV energy scales will only start to see a significant shift in their phases ($\mathcal{O}(1)$) if they are produced around or  below  the distances of the order $r_S \sim 10^4$ km, $r_S \sim 10^3 $ km, and $r_S \sim 10^2 $ km,  respectively.
     \item Therefore, those neutrinos produced with an energy scale of $E_{loc}\sim$ KeV  will be significantly affected by the rotation of the Sun up to the distance of the order  $r_S \sim 10^4$ km from the Sun. In contrast, those neutrinos that are produced with the energy scale of $E_{loc}\sim \mathcal{O}(1)$MeV and $\mathcal{O}(10)$   will only sense the rotational effect significantly up to the distance of the orders $r_S \sim 10^3$ km and $r_S \sim 10^2$ km, respectively.
    \item   For example, solar neutrinos whose production peaked around the KeV scale are produced in  the  $^7$Be, $^{15}$O, $^{13}$N, and $^{17}$F  nuclear reactions~\cite{bahcall200610}.
    \item  The flux of the neutrinos, which are mentioned above, peaks around $r_S\sim 10^4$ km~\cite{bahcall200610}, where the rotational effects on the neutrino phase remain significant for the energy scale of $E_{loc}\sim$ KeV (Eq. \ref{Sun-Earth_table_for_energy_scale})
\end{itemize}
 Hence, to demonstrate the rotational effects on the flavor flux, we,  in the following plots, take neutrinos that are produced around $r_S\sim 10^4$ km and with  an energy scale of $E_{loc}\sim$ KeV.
 In plots that are shown in Fig. \ref{fig:Suneathy0103}, the local energy of neutrinos, the  local tangential momentum of neutrinos, and the  location of the detector are taken as $E_{loc} = 0.1$ MeV, $ p^{T}_{loc}= 0.4 \, E_{loc}$ MeV/c and $r_D =10^{8}$ km,  respectively, while $r_S$ is changed from $10^4$ km to $0.5 +10^4 $ km. 
The solid black curve, dashed black curve, and black dotted curve in the Fig. \ref{fig:Suneathy0103} are plotted for the different spin parameters $y\equiv Jc/GM^2$ and  their values are $y=0.15$, $y=0.25$, and $y=0.35$, respectively\footnote{For reference, the averaged rotational parameter of the Sun is around $y\sim 0.2$.}.

\begin{itemize}
    \item In the Fig. \ref{fig:Suneathy0103}, we again see a significant relative change in the survival probabilities ($\Delta R_{ee}\%$) due to the rotation of  the gravitational source, and one can clearly see the distinction  between the black solid curve of  $y= 0.15$, black dashed curve of $y=0.25$, and blacked dotted curve of $y=0.35$, even for the  moderate rotations. 
\item We again see that there are the locations of the neutrino source where the relative change in the survival probability of the $e$ flavor is positive (from $r_S\, =0.2 \, [+10^4]$ km to $0.5 \, [+10^4]$ km). On the other hand, there are locations where the relative change in the survival probability of the $e$ flavor is negative (approximately from $r_S\, =0.05 \, [+10^4]$ km) for all the values of $y$.
\end{itemize}

Using these setups as examples, we have shown that the rotation of the gravitational source can make a difference in the transition probabilities of neutrinos when the neutrino source is located near the gravitational body but still in the weak-gravity limit. Therefore, all concerned parameters, including the rotation of the gravitational source, must be assessed individually in the neutrino oscillations while interpreting the data of neutrino flavor flux.

Now, in the numerical section,  we have used the neutrino phase (Eq \ref{Phase}), which only takes account of the leading order terms in $GM$ and $GJ$. So to estimate the error in calculating the neutrino phase, we take the ratio of the next cross term $\cal{O}$($G^2JM$) in the phase (Eq.\ref{Eq:tidelphasenextorder}, Appendix Eq. \ref{AppEq:phasenextorder})
\begin{equation}\label{Term2} 
\boldsymbol{II}= -\frac{2 G^2 J M E_0^2}{L^2} \left[\frac{L}{E_0} \left(\frac{\left(3r_D^2-4\frac{L^2}{E_0^2}\right)}{\left(r_D^2-\frac{L^2}{E_0^2}\right)^{3/2}}-\frac{\left(3r_S^2-4\frac{L^2}{E_0^2}\right)}{\left(r_S^2-\frac{L^2}{E_0^2}\right)^{3/2}}\right)- 
3\left(\sin^{-1}\left(\frac{L}{r_DE_0}\right)- \sin^{-1}\left(\frac{L}{r_SE_0}\right)\right) \right],
\end{equation}
and $\cal{O}$($GM$, $GJ$) terms 
\begin{equation}\label{Term1}
\boldsymbol{I}= \left(GM-\frac{2GJE_0}{L}\right)\left[\frac{r_D}{\sqrt{r_D^2-\left(\frac{L}{E_0}\right)^2}}-\frac{r_S}{\sqrt{r_S^2-\left(\frac{L}{E_0}\right)^2}}\right],
\end{equation}
i.e., the ratio $|\boldsymbol{II}/\boldsymbol{I}|$.
Now, for the typical parameters used for the numerical calculations,  e.g.,  $E_{loc} =10$ MeV, $p^{T}_{loc}c/E_{loc} = 0.1$,  $R_s=30$ km, $y=0.25$, $r_S = 10^5$ km and $r_D= 10^{11} $ km, we get $|\boldsymbol{II}/\boldsymbol{I}| \sim   10^{-10} $, which is a negligible number. For the $R_s=3$ km case, we have $E_{loc} =1$ MeV, $p^{T}_{loc}c/E_{loc} = 0.4$, $y=0.15$, $r_S\sim 10^6$ km and $r_D= 10^{13} $ km,  we get $|\boldsymbol{II}/\boldsymbol{I}| \sim   10^{-14} $, which is also a negligible number. And for the $R_s=3$ km case with $E_{loc} =0.1$ MeV, $p^{T}_{loc}c/E_{loc} = 0.4$, $y=0.15$, $r_S\sim 10^4$ km and $r_D= 10^{8} $ km,  we get $|\boldsymbol{II}/\boldsymbol{I}| \sim   10^{-10} $, which is again a negligible number. 
Hence, ignoring higher terms is justifiable in the presented numerical plots.
\\ 

\subsection{Decoherence Effects}\label{decohsubsec}
Up to now, we have considered neutrino oscillations in the asymptotic Kerr spacetime, assuming that neutrinos are produced as plane waves by a point-like neutrino source. However, in a real-world scenario, neutrinos are produced by a finite-sized extended neutrino source and detected by a finite-sized neutrino detector. These situations can be incorporated by using wave packet formalism~\cite{akhmedov2009paradoxes}.
Further,  wave packet formalism also introduces a new length scale $L_D$, after which neutrino oscillations cease to exist~\cite{Giunti:1991ca, giunti1998neutrinos, giunti1998coherence, grimus1999neutrino, akhmedov2009paradoxes,giunti2004coherence}. This new length scale is known as decoherence length. For the Gaussian profile, the decoherence length is given by the following relation~\cite{swami2021aspects}.
\begin{equation}\label{decohercondition}
 D_{21} = \frac{\Bar{\sigma}^2}{2}\left(|\vec{X}_{ 2}|^2-|\vec{X}_{ 1}|^2\right) \geq 1,
\end{equation}
where $\vec{X}_{i}=\partial_{\vec{p}}\Phi_{i}\left(\vec{p}=\vec{p}_{i}^{S}\right)$, $\bar{\sigma}^{2}=\sigma_{D}^{2} \sigma_{S}^{2} /\left(\sigma_{D}^{2}+\sigma_{S}^{2}\right)$ and $ D_{21}$ is the decoherence factor. Here, $\vec{p}_{i}^{S}$ is  the peak momentum of the neutrino in the local stationary frame, whereas $\sigma_S$ and $\sigma_D$ are the standard deviations in the momentum space for the neutrino and the detector wave profiles\footnote{The uncertainty  in the position space  of the produced neutrino $\sigma_x$ is  related to the uncertainty in the momentum space $\sigma_p$ as $\sigma_x \geq \hbar/2\sigma_p $, where $\hbar$ is the reduced Planck's constant.}. 
Further, the peak momentum of the detector profile $\vec{p}^D_i$  and the source $\vec{p}^S_i$ are assumed to be equal in the local frame, i.e.,  $\vec{p}^D_i= \vec{p}^S_i = \vec{p}^S$ .  Now, to estimate decoherence length, the neutrino phase  (Eq. \ref{properphaseLnotzero}) for a large proper spatial distance between the neutrino source and detector can be approximated as
\begin{equation}
    \Phi_i= \frac{m_i^2}{2E_0}L_p\Bigg( 1 - \frac{GM}{L_p}\Bigg(\tanh^{-1}\sqrt{1-\left(\frac{L}{E_0 r_D}\right)^2}-\tanh^{-1}\sqrt{1-\left(\frac{L}{E_0 r_S}\right)^2}\Bigg)\Bigg) \approx \frac{m_i^2}{2E_0}L_p.
\end{equation}
Then, using Eq. (\ref{E0/L0localequation},\ref{decohercondition}),  the proper spatial distance ($L_D$), in the leading correction to the Schwarzschild spacetime, which neutrinos will travel before they decohere, is given by 
\begin{equation}\label{Decohleng}
L_D \gtrapprox 2\sqrt{2}\left(\sqrt{B(r_S)}+\frac{R_S^2y\,p^{T}_{loc}}{r_S^2\,E_{loc}}\right) \frac{E_{\mathrm{loc}}^{2}}{\bar{\sigma} \sqrt{m_{2}^{4}-m_{1}^{4}}}.
\end{equation}
We see the coherence distance $L_D$ is also sensitive to the     neutrino tangential momentum $p^{T}_{loc}$ (equivalently $L$) and the gravitational source rotation due to the frame-dragging, as previously noted. Now, for the  parameters, $p^{T}_{loc}/E_{loc}=0.1$, $E_{loc}= 10$ MeV, $R_s =30$ km, $y=0.25$  and $r_S= 10^5$ km, and using the normal hierarchy of neutrino masses and taking the smallest neutrino mass to be zero ($\Delta m_{21}^2 = 7.42 \times 10^{-5}~{\rm eV}^2$ and $m_1^2=0$);  the  decoherance length $L_D$ for  $\sigma_D / E_{loc} = \sigma_S / E_{loc}  \approx 10^{-13} $  is of order $L_D\approx 10^{15}$ km. %
Whereas, for the  $R_s=3$ km, $p^{T}_{loc}/E_{loc}=0.4$, $E_{loc}= 1$ MeV,  $y=0.4$ km and $r_S\sim  10^6$ km,  the  decoherance length $L_D$ for  $\sigma_D / E_{loc} = \sigma_S / E_{loc}  \approx 10^{-13} $  is of order $L_D\approx 10^{14}$ km. 
Finally, for the  $R_s=3$ km, $p^{T}_{loc}/E_{loc}=0.4$, $E_{loc}= 0.1$ MeV,  $y=0.4$ km and $r_S\sim  10^4$ km,  the  decoherance length $L_D$ for  $\sigma_D / E_{loc} = \sigma_S / E_{loc}  \approx 10^{-13} $  is of order $L_D\approx 10^{13}$ km.
Hence, after this characteristic length $L_D$ between the source and the detector, all the neutrino oscillations effects will be washed away. Therefore, one requires precise information of the neutrino and the detector profile to estimate the observability of these effects in the neutrino oscillation experiments.

Further, the neutrino production and detection processes using wavepackets have been appropriately discussed in~\cite{ Kersten:2013fba,kersten2016decoherence,  porto2021coherence,akhmedov2017collective}. It has been noted that one can always restore the oscillation phenomenon at the detector if one can resolve the energy of the detector to the peak energy of the Gaussian wavepacket of the neutrinos. 
For a general scenario of neutrino production in stars/supernovas, one requires a precise model for the source of neutrino production to know whether or not the oscillation pattern gets washed away due to the collective production mechanism.
On the other hand, knowing the flavor flux of neutrinos precisely can provide information about the nature of the neutrino source.  
\\
\section{ Neutrino weak gravity lensing in the asymptotic Kerr spacetime}\label{sec:5}

In this section, we study the lensing of neutrino in the asymptotic Kerr spacetime under the weak-gravity limit. The phase of the lensed neutrino in the asymptotic Kerr spacetime, under weak gravity limit, is found to be, see Appendix \ref{App: lensing},
\begin{equation}\label{neutriphaselen}
    \Phi_i^m = \frac{m_i^2}{2E_0}\left(r_S+r_D\right)\left(1-\frac{b^2_m}{2r_Sr_D} +\frac{2GM}{r_S+r_D}- \frac{4GJ}{b_m(r_S+r_D)}\right).
\end{equation}
where $b_m$ is the impact parameter, which is  now constrained by the geometrical parameters and geometric parameters of the spacetime ($r_s, r_D, GM, GJ$),  and on the deflection angle $\delta$ of the neutrino in the weak gravity. From Eq. \ref{neutriphaselen}, we note the followings: 
\begin{itemize}
    \item  The rotational term $4GJ/b_m(r_S+r_D)$  in the neutrino phase is subdominant term. This can be seen from the ratio of $|4GJ/b_m(r_S+r_D)|$ and  $2GM/(r_S+r_D)$, which is equals to   $y R_s/b_m$.   Further,  $y$ is bounded by the naked singularity condition, i.e., $y<1$. Therefore, in the weak gravity limit, we conclude following
    \begin{equation}\label{compareterms}
    \left|\frac{4GJ}{b_m(r_S+r_D)}\right|<<\frac{2GM}{(r_S+r_D)}\implies \left|\frac{y R_s}{b_m }\right|<< 1.
    \end{equation}
    \item Further, the impact parameter $b_m$ can be shown  to be equal to the deflection angle $\delta$ in the weak gravity limit  as $\delta \simeq -4GM/b + \mathcal{O}\left(GJ/b^2\right)$~\cite{iyer2009light, kraniotis2004precise}. 
Therefore, we see $\delta$ will be modified in the order $GJ/b^2$, which is again a very suppressed term  as $\left|y R_s/2 b\right|<<1$.    Hence, effectively $\delta$ remains the same as in the weak gravity limit of Schwarzschild spacetime. Consequently, the impact parameter would not significantly modify the neutrino phase due to the rotation of gravitational source in the weak gravity limit. 
\end{itemize}
Finally, the leading order expression (orders in the $y\equiv J/GM^2$, and $R_s\equiv2GM$) of the shift in the neutrino phase ($\Delta\Phi_{shilft}^m \equiv |\Phi^m(J\neq0) - \Phi^m(J=0)|\, $) for the $i^{th}$ neutrino mass due to the rotating mass along the path $m$ can be found by plugging the $E_0$ expression (Eq. \ref{E0/L0localequation}) in the neutrino phase. The expression is given as
\begin{equation}\label{lensingshift}
\Delta\Phi_{shift}^m  \simeq  \frac{m_i^2 R_s^2\, y\,(r_D+r_S)}{4\,b_m\,E_{loc}}\left[ \frac{2}{r_D+r_S} +\frac{b_m^2}{r_S^3 }\,\left( 1-\frac{b_m^2}{r_Sr_D}\right)\right].
\end{equation}
From Eq. \ref{lensingshift},  one can find an energy scale of neutrino which can give $\mathcal{O}$ ($1$) in the shift of the neutrino phase due to the rotation of the gravitational source, which is given as
\begin{equation}\label{energyscalelens}
E_{loc}  \sim  \frac{m_i^2 R_s^2\, y\,(r_D+r_S)}{4\,b_m}\left[ \frac{2}{r_D+r_S} +\frac{b_m^2}{r_S^3 }\,\left( 1-\frac{b_m^2}{r_Sr_D}\right)\right].
\end{equation}
Now, for the numerical comparison, we take typical parameters used in the Schwarzschild neutrino lensing paper for the Sun and the Earth system~\cite{swami2020signature}, i.e.,  $r_D \sim 10^{8}$ km, and $r_S\sim10^{13}$ km\footnote{To analyze the lensing of relativistic neutrino for the small values of $r_S\sim 10^6$, one needs to study the lensing of neutrino in the strong-gravity limit. However, the study of neutrino lensing in the strong-gravity region is beyond the scope of this paper. Therefore, in the presented example, we take $r_S\sim 10^{13}$ km instead of $r_S\sim 10^6$ km. } (roughly the distance between the Alpha Centauri and Solar system, for instance) while the impact parameter is of order $|b_m|\sim \sqrt{8R_sr_D}$~\cite{swami2020signature}.
  The mass and energy of neutrino  are taken of  the orders   $m_i^2 \sim 10^{-3}$ eV$^2$, and $E_{loc}\sim1$ MeV.  Then, for any $y<1$,  we have  $|y R_s/b_m|\sim10^{-5}<<1$ and $\Delta\phi_{shift}^m \sim 10^{-4}$.  Hence, the phenomenological results of the neutrino lensing in the MeV energy scale would remain same even after accounting for the rotation of the Sun. 
However, from Eq. \ref{energyscalelens}, we find that for $r_S\sim10^{13}$ km and $r_D\sim10^8$ km,   the energy scale ($E_{loc}$) that is required for $ \Delta\Phi_{shift}^m \sim$ $ \mathcal{O}$ ($1$)  is  $E_{loc} \sim  $ sub-KeV. Although we find insignificant deviation in the neutrino phase due to the rotation of the gravitational source for the $E_{loc} \sim$ MeV \cite{swami2020signature}; there  exists an energy scale $E_{loc} \sim  $ sub-KeV,  where one may find significant deviation in the neutrino phase due to the rotation of the gravitational source.
Further, the Eqs. (\ref{lensingshift}, \ref{energyscalelens}) can be used to estimate the shift in the lensing phase of neutrino for the $i^{th}$ mass and the  $m^{th}$  path of neutrino due to the rotation of the gravitational source for different astrophysical setups.  For supermassive black holes or galaxy clusters, one may find significant deviations in the  lensing phase of neutrino even for the higher scale of neutrino energies $E_{loc}\sim$ MeV.
From Eq. \ref{lensingshift}, we also note that the rotational effect can also be made comparable with the gravitational mass effect if the naked singularity condition that bounds the Kerr rotation is violated arbitrarily, i.e., if $y$ is chosen such that $\Delta\Phi_{shift}^m \sim$ $\mathcal{O}$($1$).
 Further, $L=0$ trajectories do not contribute to weak gravity lensing even if they have non-zero $dr/d\phi$. This is because these trajectories, even if started in the weak gravity region, will not remain in the weak gravity region, see Appendix [\ref{App:L0notweaklensing}]. 
\\
Here, we point out that the  lensing of neutrino is quite different in the Kerr spacetime compared to the Schwarzschild spacetime  if we consider all the neutrino trajectory that connects the neutrino source and the detector.  In the Kerr spacetime, particles can have non-planar trajectories due to frame dragging (or broken spherical symmetry), i.e., particles can now change planes during their journey.
 These non-planar trajectories of particles can now reach the detector, which was impossible in the Schwarzschild spacetime (where particles can take only planar trajectories) and can change the oscillation probability.
Hence, one must account for all neutrino trajectories in the neutrino lensing in the broken spherical symmetry spacetimes, such as highly asymmetric mass distribution spaces and  lensing in the strong gravity regions of  the Kerr spacetime, where non-planar trajectories are expected to play a significant role.

\section{Conclusion}\label{sec:6}

Neutrino oscillation has been studied in the literature under various gravitational setups~\cite{dvornikov2019neutrino, stuttard2020neutrino,  sorge2007neutrino,lambiase2005neutrino,mastrototaro2021neutrino, lambiase2005pulsar, blasone2020neutrino, blasone2020beta, ahluwalia2015neutrino,lambiase2020effects,lambiase2020grbs,lambiase2021neutrino,buoninfante2020neutrino, dixit2019quantum,mukhopadhyay2007gravity, marletto2018quantum, motie2012high, lambiase2001neutrino, sprenger2011neutrino,chang1999possible,
ahluwalia1996gravitationally,ahluwalia1996interpretation, grossman1997flavor, bhattacharya1999gravitationally, luongo2011neutrino, geralico2012neutrino,koutsoumbas2020neutrino, chakrabarty2022effects, khalifeh2021distinguishing, khalifeh2021using, swami2020signature, swami2021aspects}. 
This study has examined neutrino flavor oscillations in the equatorial plane for the following cases in the rotating spacetime ($1$) when neutrinos are produced and detected on the same side of the gravitational source, i.e., non-lensed neutrino, and ($2$) when the neutrinos are lensed by the gravitational source in the weak gravity region.
Using the asymptotic form of the Kerr metric, we showed that the neutrino phase gets non-trivially modified due to the rotation of the gravitational source.  

For the lensed neutrinos, we find that the rotation of the gravitational source  does not substantially modify previously noted neutrino lensing results of the Schwarzschild spacetime~\cite{swami2020signature}.   We estimate the size of the modification in the neutrino phase for the Sun-Earth system.  We find that the shift in the neutrino phase $\Delta\Phi_{shift}^m$ (for $E_{loc}\sim $ MeV) due to the rotation of Sun is of $\cal{O}$ ($10^{-4}$) when the neutrino source is located at $r_S\sim10^{13}$ km (approximate distance between the Alpha Centauri and Solar system). 
However, there  exists an energy scale of neutrinos where the effects of neutrino lensing can be made significant, e.g., for the Sun-Earth case, and for the energy scale of  $E_{loc}\sim  $ sub-KeV, one  gets $\mathcal{O}$ ($1$) shift in  the neutrino phase for the $i^{th}$  neutrino mass ($m_i^2 \sim 10^{-3} $ eV$^2$) due to the rotation of the gravitational source. Further,  Eqs. (\ref{lensingshift}, \ref{energyscalelens}) are general enough for the lensing in the weak-gravity region, and one can use Eqs. (\ref{lensingshift}, \ref{energyscalelens})  to estimate the order of shift in the lensed neutrino phase due to the rotation of the gravitational source for different astrophysical objects like supermassive black holes and galaxy clusters.  In supermassive black holes or galaxy clusters, one may find significant deviations in the  lensing phase of neutrino even for the higher scale of neutrino energies $E_{loc}\sim$ MeV.
Further, we noted that the rotational effect of the gravitational source in the lensing phase of  neutrino could also be made comparable with the gravitational mass effect in the weak gravity neutrino lensing if the naked singularity condition that bounds the Kerr rotation is arbitrarily violated such that the shift in neutrino phase due to rotation of gravitational source becomes of $\mathcal{O}$ ($1$).
 This can be used as a potential discriminator between black holes and naked singularity.
\\
In contrast, for the neutrinos, which are non-lensed and emitted non-radially near the black hole (the neutrino source is still located in the weak-gravity region),  we find a \textit{significant deviation} in the neutrino flavor transition probability due to the rotation of the gravitational source when compared to the neutrino flavor transition probability in the Schwarzschild spacetime.
We explicitly demonstrate the qualitative effects of the gravitational source rotation on the neutrino flavor transition probabilities using numerical examples for a realistic three neutrino flavor case.
 It is seen that, unlike in the Schwarzschild spacetime, neutrino flavor transition probability is very sensitive to the relative gravitational source rotation and neutrino angular momentum directions. This is because, in one case, neutrinos have to traverse along with the rotation of the spacetime and, in the other case, neutrinos have to traverse against the rotation of the spacetime due to the frame-dragging induced by the rotation of the gravitational source. Therefore, in these two cases, one gets different phase of the neutrino.
Further, we find that neutrinos that are produced with the same local 4-momentum have a quite distinct appearance probability of neutrino flavors in the rotating and Schwarzschild spacetime.
Furthermore, we give numerical examples using the Sun-sized gravitational source ($R_s \sim3$ km), and the location of the neutrino source $r_D$ is taken as $r_D\sim10^{13}$ km, which corresponds to the distance between the  Alpha Centauri and the Solar system and $r_D\sim 10^8$ km, which corresponds to the distance between the Sun and the Earth.
We show that, in both cases, even a small change in the rotation of the gravitational object (i.e., $y=0.15$)  gives a significant relative change in the survival probability of electron ($e$) flavor of  neutrino ($\Delta R_{ee}\%$).
 We find that the cases when $y=0.4$ and $y=0.35$, the relative change in the survival probability of electron ($e$) flavor of neutrino for the particular neutrino source location can be as significant as $300\%$. 
 Therefore, for a neutrino source that is located around a  region ($r_S$) where $\Delta R_{ee}(\%)$ is positive,  the detector would register a net jump in the survival events of the electron ($e$) flavor of neutrino.
In contrast, for a neutrino source that is located around a region where $\Delta R_{ee}\%$  is negative, the detector would register a net drop in the survival events of the electron ($e$) flavor of neutrino. 
Also,  we see a clear difference in the curves of $\Delta R_{ee}(\%)$  for the different values of the spin parameter $y$. 
Hence, the rotation of the gravitational source can change the survival or appearance event counts as registered by the detector.  
Further,  one can readily  use  Eq. (\ref{onesideshift}) for the one-sided neutrino propagation $r_D/r_S<<1$  to estimate the order of shift in the neutrino phase due to the rotation of the gravitational source for different astrophysical setups.
We note that the presented study is restricted to the weak-gravity region; a complete study would include MSW effects~\cite{wolfenstein1978matter,mikheev1985resonance} inside the stars and modifications from the internal metric of the star that will further modify the oscillation length of neutrino as the case even for the Schwarzschild analysis, and including these effects is beyond the scope of the paper and should be taken elsewhere.
However, we also note that the MSW effects are not always prominent, even in the matter region of the neutrino source. Therefore, the presented results would get minimal corrections from MSW effects in those cases. 
For example, in a case, when the matter interaction potential $V$ is such that $V<<\Delta m_{ij}^2/2E_{loc}$, one gets a small contribution from the matter interaction term in the oscillatory phase of neutrino~\cite{fantini2018formalism} and therefore can be ignored as the fractional contribution of  MSW effects in the neutrino phase will fall as neutrinos travel astronomical or cosmological distances to reach the detector. 
 These results have implications for the study of neutrinos, which have astrophysical origins like supernovas, as they travel a large distance to reach the earth; also, in the study of cosmic neutrino background, which presumably passes through many galaxies and supermassive black holes before reaching the earth. 
Therefore, we conclude that the rotational effects of the gravitational source can not be discarded while understanding the data of neutrino flux from the astrophysical source, as in these cases, the rotation of the gravitational source becomes an important parameter in the analysis of neutrino flux.

Apart from including the effects of rotating gravitation sources on the neutrinos, which travel for large distances, one may also include the effects on the oscillation phase due to quantum gravity. These terms may also come alive at large distances, like in the case of the effects of rotating gravitational sources on the oscillation phase.
Quantum gravity effects can be incorporated into the neutrino phase by including corrections in the metric tensor due to quantum gravity. For example, in~\cite{bjerrum2003quantum,calmet2019quantum}, the quantum corrections to the Kerr and Schwarzschild metric and a star metric are examined, respectively.
Due to the change in the arrival time (or equivalently, the proper time spent by the neutrino) because of the quantum corrections, the change in the neutrino phase, which is proportional to the proper time (see Appendix \ref{app:nullvsmassivetrajectory}), will acquire a different value, and that can also change the flavor flux of neutrinos registered by the detector.
\\

\section{Acknowledgement}
HS thanks Kinjalk Lochan and Ketan M. Patel for the invaluable comments and suggestions during the  preparation of the manuscript and for careful readings of the manuscript. HS would also like to thank Council of Scientific \& Industrial Research (CSIR), India, for the financial support through research fellowship award no. 514856. This research is also partially supported by the Department of Science and Technology, India, through a project grant under DST/INSPIRE/04/2016/000571.
\appendix

\section{Neutrino massive trajectory vs null trajectory approximation}\label{app:nullvsmassivetrajectory}
Ignoring spin structure of neutrino, the relevant phase for the $i^{th}$ massive neutrino is given by 
\begin{equation}\label{phaseApp}
    \Phi_i = \int p_\mu^i dx^\mu.
\end{equation}
However, using standard neutrino phase under the null trajectory approximation for a relativistic neutrino is shown to be~\cite{cardall1997neutrino}
\be\label{nullphase}
\Phi_i^{null}= \int p^{i}_\mu dx^{\mu}|_{null} = \int p^{i}_\mu p^{\mu}_{null} d\lambda =\frac{m_i^2}{2}\int d\lambda= \frac{m_i^2}{2}\lambda,
\ee
where  $\lambda$ is affine parameter along the null geodesic. 
Whereas if we take a massive trajectory for the neutrino, we get from Eq. \ref{phaseApp}
\begin{equation}\label{phasemassive}
      \Phi_i^{massive} = \int p_\mu^i dx^\mu|_{massive} = \int p_\mu^i \frac{dx^\mu}{d\tau_i}d\tau_i,
\end{equation}
where  $\tau_i$ is the proper time along the neutrino massive geodesic.
Now, using $ p_\mu = m \, dx^\mu/d\tau$ and on mass shell condition $p_\mu \, p^\mu = m^2$, we get
\begin{equation}\label{masstauphase}
      \Phi_i^{massive} = m_i\, \int d\tau_i\,.
\end{equation}
Now, we can parameterize the geodesic with any one of the coordinates $t(\tau)$, $r(\tau)$, $\theta(\tau)$ or $\phi(\tau)$ provided it has a non-zero first derivative w.r.t. $\tau$ in the region. Let us parameterize geodesics, for the sake of the calculation, with radial coordinate $r$. Then, we can rewrite Eq. (\ref{nullphase}, \ref{phasemassive}) as
\begin{align}
     \Phi_i^{null} = \frac{m_i^2}{2}\, \int \frac{dr}{(p^r_i)^{null}}\,, \nonumber\\ \Phi_i^{massive} = m_i^2\, \int \frac{dr}{(p^r_i)^{massive}}\,.
\end{align}
 Maintaining neutrino phase only in the leading power of $m$, we can expand $(p^r_i)^{massive}$ as 
\begin{equation}\label{radialmomentumexpansion}
      (p^r_i)^{massive} =(p^r)^{null} + corrections\,  in \, m_i,
\end{equation}
and putting it in $\Phi_i^{massive}$, we get
\begin{equation}\label{masstauphaseradialprameterize}
      \Phi_i^{massive} = m_i^2\, \int \frac{dr}{(p^r)^{null}} +  corrections\,   of \, higher\, order \, in\, m_i \,.
\end{equation}
which is twice of the phase we get from null trajectory approximation, i.e., 2 $\Phi_i^{null} = \Phi_i^{massive}= m_i \tau_i$, in the leading order of neutrino mass $m_i$.

\section{Critical geodesic solution for $\theta(\tau)$ in $\theta = \pi/2$ plane}
Geodesic equation for $p_\theta$ is given as
\begin{equation}\label{theta geodesic}
    m\frac{d p_\theta}{d\tau}= \frac{1}{2}\, \partial_\theta g_{\mu\nu}\, p^\mu\, p^\nu.
\end{equation}
For the following metric, \be ds^2 = g_{tt}\, dt^2 + 2g_{t\phi} \, dt \,d\phi+g_{rr}\, dr^2+ g_{\theta\theta}\, d\theta^2 + g_{\phi\phi}\, d\phi^2\,,\ee
where $g_{tt}= 1-R_s/r$, $g_{rr}= -1/g_{tt}$, $g_{\theta\theta}=-r^2$, $g_{\phi\phi}=-r^2\sin^2\theta$, and $g_{t\phi}=2GJ\sin^2\theta/r$, \\
the Eq. \ref{theta geodesic}, after some manipulation,  can also be written as
\begin{equation}\label{Eq. theta geodesic phi parameter}
    m\frac{dp_\theta}{d\tau}= p^\phi(g_{\phi\phi}\, p^\phi +2\, g_{t\phi}\, p^t)\cot\theta.
\end{equation}
Now, using  conserved quantities $p_t$ and $p_\phi$, i.e., $p_{\phi} = g_{\phi\phi}\,p^\phi +g_{\phi t}\,p^t$ ,
$p_{t} = g_{\phi t}\,p^\phi +g_{t t}\,p^t$, and $p_\theta = m\, g_{\theta\theta}\,d\theta/d\tau$, we can rewrite Eq. \ref{Eq. theta geodesic phi parameter} as
\begin{equation}\label{Eq. theta geodesic phi parameter3}
m^2 g_{\theta\theta} \,\frac{d^2\theta}{d \tau^2}+ m^2 \frac{dg_{\theta\theta}}{d\tau} \frac{d\theta}{d \tau}- \left(p_\phi+ \frac{g_{\phi t}(g_{\phi\phi}p_t-g_{\phi t}p_\phi) }{g_{\phi \phi}g_{tt}-g_{\phi t}^2}\right)\left(\frac{g_{tt}p_\phi-g_{\phi t}p_t }{g_{\phi \phi}g_{tt}-g_{\phi t}^2}\right)\cot\theta=0.
\end{equation}

As all coefficients in above D.E.  are well defined at $\theta= \pi/2$, we  see that the above D.E. for the initial values  $\theta =\pi/2$ and  $d\theta/d\tau =0$ has a critical solution, i.e., for these initial values we get $d^2\theta/d\tau^2 =0$. So, the D.E. has a solution $\theta(\tau) =\pi/2 = constant$  along the trajectory. Hence, all the particles which are produced with initial values $\theta =\pi/2$ and  $d\theta/d\tau =0$  remain in the $\theta =\pi/2$ plane throughout its journey.

\section{Neutrino phase change in asymptotic Kerr geometry under null trajectory approximation}\label{neutrinophase cal}

Let us parametrize geodesic with radial coordinate; then we can rewrite neutrino phase Eq. \ref{nullphase} as

\be\label{drphase}
\Phi_i = \frac{m_i^2}{2}\int \frac{1}{p^{\phi}}\frac{d\phi}{dr}dr
\ee
Using null condition $(p_{\mu}p^{\mu})_{null}=0$, we get
\be\label{Appdedphi}
\frac{dr}{d\phi}= \pm\sqrt{\left(B(r)\frac{p^t}{p^\phi}\right)^2+4B(r)\frac{GJ}{r}\frac{p^t}{p^\phi}-B(r)r^2}
\ee
Now, two conserved quantities $p_t=E_0$ and $p_\phi\equiv-L$ along the geodesic are related to $p^t$ and $p^\phi$ by the following relation
\be\label{App(pt)}
p^{t} = \frac{E_0 r^2 -h(r) L}{h(r)^2+r^2B(r)},
\ee
and
\be\label{App(p(phi))}
p^{\phi} = \frac{E_0h(r)+B(r)L}{h(r)^2+r^2B(r)},
\ee
where $h(r)= 2GJ/r$, $B(r)=1-2GM/r$.

\subsection{Neutrino produced with non-zero angular  momentum ($L\neq0$) (The source and detector lie on the same side of the gravitational mass)}
For $L\neq0$, using Eq. (\ref{Appdedphi}, \ref{App(p(phi))}, \ref{App(pt)}) in Eq. \ref{drphase}, we get for $r_D>r_S$ 
\begin{equation}\label{intphaseL}
\Phi_i = \frac{m_i^2}{2}\int_{r_D}^{r_S} \frac{h(r)^2+r^2B(r)}{\left(h(r)E_0+LB(r)\right)\sqrt{ \left(B(r)\frac{E_0 r^2 -h(r)L}{h(r)E_0+LB(r) }\right)^2+ 2h(r) B(r)\frac{E_0 r^2 -h(r)L}{h(r)E_0+LB(r)} -r^2 B(r)}}dr.
\end{equation}
Now, expanding $\Phi_i$ keeping the leading order of $GM/r$, $GJ/r^2$ and the cross term $G^2JM/r^3$  under weak field limit, we find
\begin{equation}
\Phi_i \simeq \frac{m_i^2}{2E_0}\int_{r_D}^{r_S}\left(\frac{r}{\sqrt{r^2-\frac{L^2}{ E_0^2}}}+ \Big(\frac{2GJ L}{E_0} -\frac{GML^2}{E_0^2}\Big)\frac{1}{\left(r^2-\frac{L^2}{E_0^2}\right)^{3/2}}
-\frac{6 G^2 J M L^3}{E_0^3}\frac{1}{r\, 
\left(r^2-\frac{L^2}{E_0^2}\right)^{5/2}}\right) dr.
\end{equation}
Integrating for condition $r>L/E_0$ we get
 \begin{align}\label{AppEq:phasenextorder} 
\Phi_i\simeq & \frac{m_i^2}{2E_0}\Bigg(\sqrt{r_D^2-\left(\frac{L}{E_0}\right)^2} -\sqrt{r_S^2-\left(\frac{L}{E_0}\right)^2} +\left(GM-\frac{2GJE_0}{L}\right)\left(\frac{r_D}{\sqrt{r_D^2-\left(\frac{L}{E_0}\right)^2}}-\frac{r_S}{\sqrt{r_S^2-\left(\frac{L}{E_0}\right)^2}}\right) 
 \nonumber \\ & -\frac{2 G^2 J M E_0}{L}  \left(\frac{\left(3r_D^2-4\frac{L^2}{E_0^2}\right)}{\left(r_D^2-\frac{L^2}{E_0^2}\right)^{3/2}}-\frac{\left(3r_S^2-4\frac{L^2}{E_0^2}\right)}{\left(r_S^2-\frac{L^2}{E_0^2}\right)^{3/2}}\right)+ 
\frac{6 G^2 J M E_0^2}{L^2}\left(\sin^{-1}\left(\frac{L}{r_DE_0}\right)- \sin^{-1}\left(\frac{L}{r_SE_0}\right)\right) \Bigg).
\end{align}

\subsection{Neutrino produced with zero angular  momentum ($L=0$)}

Now, similarly for $L= 0$,  we get for $r_D>r_S$
\be \label{AppFramegragging}
\frac{d \phi}{d r} = \frac{2GJ}{\sqrt{\left(1-\frac{2GM}{r}\right)\left(r^6-2GMr^5+4G^2r^2J^2\right)}}
.
\ee
Using Eq. \ref{AppFramegragging} and  Eq. \ref{App(p(phi))} in Eq. \ref{drphase} and integrating from $r_S$ to $r_D$, we get phase in the leading order after ignoring order ${\cal{O}}\left(G^4M^2J^2/r^4\right) $ 
\begin{align}
\Phi_i \simeq & \frac{m_i^2}{2E_0}\left[\left(r_D-r_S\right)-\frac{2G^2J^2}{3}\left(\frac{1}{r_D^3}-\frac{1}{r_S^3}\right)-G^3MJ^2\left(\frac{1}{r_D^4}-\frac{1}{r_S^4}\right) +{\cal{O}}\left(\frac{G^4M^2J^2}{r_{S,D}^5}\right)\right].
\end{align}


\subsection{Neutrino phase change in the weak gravitational lensing in the $\theta=\pi/2$ plane}\label{App: lensing}
In the weak gravitational lensing, one usually defines impact parameter $b$ in the asymptotic observer frame as $b\equiv L/E_0$, where $L$ and $E_0$ is the angular momentum and energy of the particle as measured by the asymptotic observer. Now, replacing $L$ and $E_0$ in term of $b$ in the Eq. \ref{intphaseL}, we get 
\begin{equation}\label{intphaselensing}
\Phi_i = \frac{m_i^2}{2E_0}\int_{r_D}^{r_S} \frac{h(r)^2+r^2B(r)}{\left(h(r)+bB(r)\right)\sqrt{ \left(B(r)\frac{ r^2 -h(r)b}{h(r)+bB(r) }\right)^2+ 2h(r) B(r)\frac{ r^2 -h(r)b}{h(r)+bB(r)} -r^2 B(r)}}dr.
\end{equation}
Now, limit of integral is  broken into two regions of neutrino propagation (1) when neutrino is proporgating towards gravitational source, and (2) when neutrino is propagating away from the gravitational source, i.e., $(r_S\rightarrow r_0 \rightarrow r_D)$,
\begin{equation}
\Phi_i^{lens} = \frac{m_i^2}{2E_0}\left(  \int_{r_0}^{r_D}-\int_{r_S}^{r_0} \right)\left[ \frac{h(r)^2+r^2B(r)}{\left(h(r)+bB(r)\right)\sqrt{ \left(B(r)\frac{ r^2 -h(r)b}{h(r)+bB(r) }\right)^2+ 2h(r) B(r)\frac{ r^2 -h(r)b}{h(r)+bB(r)} -r^2 B(r)}}\right]dr. 
\end{equation}
The $-ve$ sign accounts for the neutrino traveling towards the gravitational source ($dr$ decreasing). Where $r_0$ is the radius of closest approach and can be found in terms of  $b$ using following equation
\begin{equation}\label{closestapproach}
    \frac{dr}{d\phi}=0=  \sqrt{\left(B(r)\frac{p^t}{p^\phi}\right)^2+4B(r)\frac{GJ}{r}\frac{p^t}{p^\phi}-B(r)r^2}
\end{equation}
 Solving Eq. \ref{closestapproach}, in the leading order of $M$ and $J$, we find
\begin{align}\label{b and r_0}
    b\simeq r_0+GM -\frac{2GJ}{r_0}\,,\nonumber \\ r_0 \simeq b - GM +\frac{2GJ}{b} \,.
\end{align}
Now plugging $b$ in terms of $r_0$ in Eq. \ref{intphaselensing}, and expanding Eq. \ref{intphaselensing} up to the leading  orders of $GM/r$ and $GJ/r^2$. We get after integration   
\begin{equation}
 \Phi_i^{lens} \simeq \frac{m_i^2}{2E_0}\left[\sqrt{r_{D}^{2}-r_{0}^{2}}+\sqrt{r_{S}^{2}-r_{0}^{2}}+\left(G M-\frac{2GJ}{r_0}\right)\left(\sqrt{\frac{r_{D}-r_{0}}{r_{D}+r_{0}}}+\sqrt{\frac{r_{S}-r_{0}}{r_{S}+r_{0}}}\right)\right].
\end{equation}
Now replacing $r_0$ in terms of $b$ from Eq. \ref{b and r_0}, we get in the weak field limit $b/r_{S,D}<<1$ (ignoring higher order in $b^2/r^2_{S,D}$,  $GM/r$ and $GJ/r^2$)
\begin{equation}
\Phi_i^{lens} \simeq \frac{m_{i}^{2}}{2 E_{0}}\left(r_{S}+r_{D}\right)\left[1-\frac{b^{2}}{2 r_{S} r_{D}}+\frac{2 G M}{r_{S}+r_{D}}-\frac{4GJ}{b\left(r_S+r_D\right)}\right]
\end{equation}

\subsubsection{$L=0$ trajectories do not contribute in the weak gravity lensing}\label{App:L0notweaklensing}
To see that neutrinos produced with $L=0$ do not contribute to the weak gravity lensing, we find radial coordinate ($r_t$) at which these trajectories turn around, i.e., when $dr/d\phi =0$, then using Eq. \ref{AppFramegragging}, we get
\begin{equation}
    \left(1-\frac{2GM}{r_t}\right)\left(r_t^6-2GMr_t^5+4G^2r_t^2J^2\right)= 0.
\end{equation}
This gives following solutions
\begin{eqnarray}
 r_t&=0, \nonumber\\  r_t& = R_s \\   r_t^4\left(1-\frac{2GM}{r_t}\right) +4 G^2 J^2& =0\nonumber
\end{eqnarray}
The first two solutions violate the weak gravity limit, while the third equation has no real solution.
 Hence, the $L=0$ case is irrelevant for the weak gravity lensing.

\bibliographystyle{unsrt}
\bibliography{bibliography.bib}

\begin{thebibliography}{10}

\bibitem{capozzi2014status}
F~Capozzi, GL~Fogli, E~Lisi, A~Marrone, D~Montanino, and A~Palazzo.
\newblock Status of three-neutrino oscillation parameters, circa 2013.
\newblock {\em Physical Review D}, 89(9):093018, 2014.

\bibitem{de2018status}
Pablo~Fern{\'a}ndez de~Salas, DV~Forero, Christoph~Andreas Ternes, Mariam
  T{\'o}rtola, and Jos{\'e}~WF Valle.
\newblock Status of neutrino oscillations 2018: 3$\sigma$ hint for normal mass
  ordering and improved cp sensitivity.
\newblock {\em Physics Letters B}, 782:633--640, 2018.

\bibitem{esteban2019global}
Ivan Esteban, Maria~Concepti{\'o}n Gonz{\'a}lez-Garc{\'\i}a, Alvaro
  Hernandez-Cabezudo, Michele Maltoni, and Thomas Schwetz.
\newblock Global analysis of three-flavour neutrino oscillations: synergies and
  tensions in the determination of $\theta$ 23, $\delta$ cp, and the mass
  ordering.
\newblock {\em Journal of High Energy Physics}, 2019(1):1--35, 2019.

\bibitem{esteban2020fate}
Ivan Esteban, Maria~Concepti{\'o}n Gonz{\'a}lez-Garc{\'\i}a, Michele Maltoni,
  Thomas Schwetz, and Albert Zhou.
\newblock The fate of hints: updated global analysis of three-flavor neutrino
  oscillations.
\newblock {\em Journal of High Energy Physics}, 2020(9):1--22, 2020.

\bibitem{dvornikov2019neutrino}
Maxim Dvornikov.
\newblock Neutrino flavor oscillations in stochastic gravitational waves.
\newblock {\em Physical Review D}, 100(9):096014, 2019.

\bibitem{stuttard2020neutrino}
Thomas Stuttard and Mikkel Jensen.
\newblock Neutrino decoherence from quantum gravitational stochastic
  perturbations.
\newblock {\em Physical Review D}, 102(11):115003, 2020.

\bibitem{mavromatos2008quantum}
Nick~E Mavromatos, Anselmo Meregaglia, Andre Rubbia, Alexander~S Sakharov, and
  Sarben Sarkar.
\newblock Quantum-gravity decoherence effects in neutrino oscillations:
  Expected constraints from cngs and j-parc.
\newblock {\em Physical Review D}, 77(5):053014, 2008.

\bibitem{alfaro2002quantum}
Jorge Alfaro, Hugo~A Morales-Tecotl, and Luis~F Urrutia.
\newblock Quantum gravity and spin-1/2 particle effective dynamics.
\newblock {\em Physical Review D}, 66(12):124006, 2002.

\bibitem{marletto2018quantum}
C~Marletto, V~Vedral, and D~Deutsch.
\newblock Quantum-gravity effects could in principle be witnessed in
  neutrino-like oscillations.
\newblock {\em New Journal of Physics}, 20(8):083011, 2018.

\bibitem{katori2006global}
Teppei Katori, V~Alan Kosteleck{\`y}, and Rex Tayloe.
\newblock Global three-parameter model for neutrino oscillations using lorentz
  violation.
\newblock {\em Physical Review D}, 74(10):105009, 2006.

\bibitem{diaz2009perturbative}
Jorge~S Diaz, V~Alan Kosteleck{\`y}, and Matthew Mewes.
\newblock Perturbative lorentz and c p t violation for neutrino and
  antineutrino oscillations.
\newblock {\em Physical Review D}, 80(7):076007, 2009.

\bibitem{antonelli2018neutrino}
Vito Antonelli, L~Miramonti, and MDC Torri.
\newblock Neutrino oscillations and lorentz invariance violation in a
  finslerian geometrical model.
\newblock {\em The European Physical Journal C}, 78(8):1--13, 2018.

\bibitem{motie2012high}
Iman Motie and She-Sheng Xue.
\newblock High energy neutrino oscillation at the presence of the lorentz
  invariance violation.
\newblock {\em International Journal of Modern Physics A}, 27(19):1250104,
  2012.

\bibitem{lambiase2001neutrino}
G~Lambiase.
\newblock Neutrino oscillations in non-inertial frames and the violation of the
  equivalence principle neutrino mixing induced by the equivalence principle
  violation.
\newblock {\em The European Physical Journal C-Particles and Fields},
  19(3):553--560, 2001.

\bibitem{sprenger2011neutrino}
Martin Sprenger, Piero Nicolini, and Marcus Bleicher.
\newblock Neutrino oscillations as a novel probe for a minimal length.
\newblock {\em Classical and Quantum Gravity}, 28(23):235019, 2011.

\bibitem{chang1999possible}
Chao-Hsi Chang, Wu-Sheng Dai, Xue-Qian Li, Yong Liu, Feng-Cai Ma, and Zhi-jian
  Tao.
\newblock Possible effects of quantum mechanics violation induced by certain
  quantum-gravity effects on neutrino oscillations.
\newblock {\em Physical Review D}, 60(3):033006, 1999.

\bibitem{sorge2007neutrino}
Francesco Sorge and Silvio Zilio.
\newblock Neutrino spin flip around a schwarzschild black hole.
\newblock {\em Classical and Quantum Gravity}, 24(10):2653, 2007.

\bibitem{lambiase2005neutrino}
Gaetano Lambiase, Giorgio Papini, Raffaele Punzi, and Gaetano Scarpetta.
\newblock Neutrino optics and oscillations in gravitational fields.
\newblock {\em Physical Review D}, 71(7):073011, 2005.

\bibitem{mastrototaro2021neutrino}
Leonardo Mastrototaro and Gaetano Lambiase.
\newblock Neutrino spin oscillations in conformally gravity coupling models and
  quintessence surrounding a black hole.
\newblock {\em arXiv preprint arXiv:2106.07665}, 2021.

\bibitem{lambiase2005pulsar}
Gaetano Lambiase.
\newblock Pulsar kicks induced by spin flavour oscillations of neutrinos in
  gravitational fields.
\newblock {\em Monthly Notices of the Royal Astronomical Society},
  362(3):867--871, 2005.

\bibitem{ahluwalia1996gravitationally}
Dharam~Vir Ahluwalia and C~Burgard.
\newblock Gravitationally induced neutrino-oscillation phases.
\newblock {\em General Relativity and Gravitation}, 28(10):1161--1170, 1996.

\bibitem{ahluwalia1996interpretation}
Dharam~Vir Ahluwalia and C~Burgard.
\newblock About the interpretation of gravitationally induced neutrino
  oscillation phases.
\newblock {\em arXiv preprint gr-qc/9606031}, 1996.

\bibitem{grossman1997flavor}
Yuval Grossman and Harry~J Lipkin.
\newblock Flavor oscillations from a spatially localized source: A simple
  general treatment.
\newblock {\em Physical Review D}, 55(5):2760, 1997.

\bibitem{bhattacharya1999gravitationally}
Tanmoy Bhattacharya, Salman Habib, and Emil Mottola.
\newblock Gravitationally induced neutrino oscillation phases in static
  spacetimes.
\newblock {\em Physical Review D}, 59(6):067301, 1999.

\bibitem{luongo2011neutrino}
Orlando Luongo and Gabriele~V. Stagno.
\newblock Neutrino oscillation at the lifshitz point.
\newblock {\em Mod. Phys. Lett. A}, 26(17):1257, 2011.

\bibitem{geralico2012neutrino}
Andrea Geralico and Orlando Luongo.
\newblock Neutrino oscillations in the field of a rotating deformed mass.
\newblock {\em Physics Letters A}, 376(15):1239--1243, 2012.

\bibitem{koutsoumbas2020neutrino}
George Koutsoumbas and Dimitrios Metaxas.
\newblock Neutrino oscillations in gravitational and cosmological backgrounds.
\newblock {\em General Relativity and Gravitation}, 52(10):1--13, 2020.

\bibitem{chakrabarty2022effects}
Hrishikesh Chakrabarty, Debasish Borah, Ahmadjon Abdujabbarov, Daniele
  Malafarina, and Bobomurat Ahmedov.
\newblock Effects of gravitational lensing on neutrino oscillation in
  $\gamma$-spacetime.
\newblock {\em The European Physical Journal C}, 82(1):1--15, 2022.

\bibitem{blasone2020neutrino}
Massimo Blasone, Gaetano Lambiase, Giuseppe~Gaetano Luciano, and Luciano
  Petruzziello.
\newblock Neutrino oscillations in unruh radiation.
\newblock {\em Physics Letters B}, 800:135083, 2020.

\bibitem{blasone2020beta}
Massimo Blasone, Gaetano Lambiase, Giuseppe~Gaetano Luciano, and Luciano
  Petruzziello.
\newblock On the $\beta$-decay of the accelerated proton and neutrino
  oscillations: a three-flavor description with cp violation.
\newblock {\em The European Physical Journal C}, 80(2):130, 2020.

\bibitem{ahluwalia2015neutrino}
Dharam~Vir Ahluwalia, Lance Labun, and Giorgio Torrieri.
\newblock Neutrino oscillations in accelerated states.
\newblock {\em arXiv preprint arXiv:1508.03091}, 2015.

\bibitem{lambiase2020effects}
Gaetano Lambiase and Leonardo Mastrototaro.
\newblock Effects of modified theories of gravity on neutrino pair annihilation
  energy deposition near neutron stars.
\newblock {\em The Astrophysical Journal}, 904(1):19, 2020.

\bibitem{lambiase2020grbs}
G~Lambiase and L~Mastrototaro.
\newblock Grbs from neutrino pair annihilation in the presence of quintessence
  surrounding a black hole.
\newblock {\em arXiv e-prints}, pages arXiv--2012, 2020.

\bibitem{lambiase2021neutrino}
G.~Lambiase and L.~Mastrototaro.
\newblock Neutrino pair annihilation $\left(v \bar{v} \rightarrow e^{-}
  e^{+}\right)$ in the presence of quintessence surrounding a black hole.
\newblock {\em The European Physical Journal C}, 81(10):932, 2021.

\bibitem{buoninfante2020neutrino}
Luca Buoninfante, Giuseppe~Gaetano Luciano, Luciano Petruzziello, and Luca
  Smaldone.
\newblock Neutrino oscillations in extended theories of gravity.
\newblock {\em Physical Review D}, 101(2):024016, 2020.

\bibitem{dixit2019quantum}
Khushboo Dixit, Javid Naikoo, Banibrata Mukhopadhyay, and Subhashish Banerjee.
\newblock Quantum correlations in neutrino oscillations in curved spacetime.
\newblock {\em Physical Review D}, 100(5):055021, 2019.

\bibitem{mukhopadhyay2007gravity}
Banibrata Mukhopadhyay.
\newblock Gravity-induced neutrino--antineutrino oscillation: Cpt and lepton
  number non-conservation under gravity.
\newblock {\em Classical and Quantum Gravity}, 24(6):1433, 2007.

\bibitem{khalifeh2021distinguishing}
Ali~Rida Khalifeh and Raul Jimenez.
\newblock Distinguishing dark energy models with neutrino oscillations.
\newblock {\em Physics of the Dark Universe}, 34:100897, 2021.

\bibitem{khalifeh2021using}
Ali~Rida Khalifeh and Raul Jimenez.
\newblock Using neutrino oscillations to measure $ h\_0$.
\newblock {\em arXiv preprint arXiv:2111.15249}, 2021.

\bibitem{swami2020signature}
Himanshu Swami, Kinjalk Lochan, and Ketan~M Patel.
\newblock Signature of neutrino mass hierarchy in gravitational lensing.
\newblock {\em Physical Review D}, 102(2):024043, 2020.

\bibitem{swami2021aspects}
Himanshu Swami, Kinjalk Lochan, and Ketan~M Patel.
\newblock Aspects of gravitational decoherence in neutrino lensing.
\newblock {\em Physical Review D}, 104(9):095007, 2021.

\bibitem{ren2010neutrino}
Jun Ren and Cheng-Min Zhang.
\newblock Neutrino oscillations in the kerr--newman spacetime.
\newblock {\em Classical and Quantum Gravity}, 27(6):065011, 2010.

\bibitem{visinelli2015neutrino}
Luca Visinelli.
\newblock Neutrino flavor oscillations in a curved space-time.
\newblock {\em General Relativity and Gravitation}, 47(5):1--17, 2015.

\bibitem{uribe2021neutrino}
Juan~David Uribe, Eduar~Antonio Becerra-Vergara, and Jorge~Armando Rueda.
\newblock Neutrino oscillations in neutrino-dominated accretion around rotating
  black holes.
\newblock {\em Universe}, 7(1):7, 2021.

\bibitem{baines2021painleve}
Joshua Baines, Thomas Berry, Alex Simpson, and Matt Visser.
\newblock Painleve$^`$-gullstrand form of the lense-thirring spacetime.
\newblock {\em Universe}, 7(4):105, 2021.

\bibitem{visser2007kerr}
Matt Visser.
\newblock The kerr spacetime: A brief introduction.
\newblock {\em arXiv preprint arXiv:0706.0622}, 2007.

\bibitem{Stoghianidis1987polar}
E.~Stoghianidis and D.~Tsoubelis.
\newblock Polar orbits in the kerr space-time.
\newblock {\em General Relativity and Gravitation}, 19:1235--1249, 1987.

\bibitem{kraniotis2004precise}
GV~Kraniotis.
\newblock Precise relativistic orbits in kerr and kerr--(anti) de sitter
  spacetimes.
\newblock {\em Classical and Quantum Gravity}, 21(19):4743, 2004.

\bibitem{iyer2009light}
Savitri~V Iyer and Edward~C Hansen.
\newblock Light's bending angle in the equatorial plane of a kerr black hole.
\newblock {\em Physical Review D}, 80(12):124023, 2009.

\bibitem{cardall1997neutrino}
Christian~Y Cardall and George~M Fuller.
\newblock Neutrino oscillations in curved spacetime: A heuristic treatment.
\newblock {\em Physical Review D}, 55(12):7960, 1997.

\bibitem{Giunti:1991ca}
C.~Giunti, C.~W. Kim, and U.~W. Lee.
\newblock {When do neutrinos really oscillate?: Quantum mechanics of neutrino
  oscillations}.
\newblock {\em Phys. Rev. D}, 44:3635--3640, 1991.

\bibitem{giunti1998neutrinos}
C~Giunti, CW~Kim, and UW~Lee.
\newblock When do neutrinos cease to oscillate?
\newblock {\em Physics Letters B}, 421(1-4):237--244, 1998.

\bibitem{giunti1998coherence}
Carlo Giunti and Chung~W Kim.
\newblock Coherence of neutrino oscillations in the wave packet approach.
\newblock {\em Physical Review D}, 58(1):017301, 1998.

\bibitem{grimus1999neutrino}
Walter Grimus, Subhendra Mohanty, and Peter Stockinger.
\newblock Neutrino oscillations and the effect of the finite lifetime of the
  neutrino source.
\newblock {\em Physical Review D}, 61(3):033001, 1999.

\bibitem{akhmedov2009paradoxes}
E~Kh Akhmedov and A~Yu Smirnov.
\newblock Paradoxes of neutrino oscillations.
\newblock {\em Physics of Atomic Nuclei}, 72(8):1363--1381, 2009.

\bibitem{giunti2004coherence}
Carlo Giunti.
\newblock Coherence and wave packets in neutrino oscillations.
\newblock {\em Foundations of Physics Letters}, 17(2):103--124, 2004.

\bibitem{chatelain2020neutrino}
Am{\'e}lie Chatelain and Maria~Cristina Volpe.
\newblock Neutrino decoherence in presence of strong gravitational fields.
\newblock {\em Physics Letters B}, 801:135150, 2020.

\bibitem{luciano2021gravitational}
Giuseppe~Gaetano Luciano and Massimo Blasone.
\newblock Gravitational effects on neutrino decoherence in the lense--thirring
  metric.
\newblock {\em Universe}, 7(11):417, 2021.

\bibitem{sadeghi2021wave}
P~Sadeghi, F~Hammad, A~Landry, and T~Martel.
\newblock Wave packet treatment of neutrino flavor oscillations in various
  spacetimes.
\newblock {\em General Relativity and Gravitation}, 53(11):1--28, 2021.

\bibitem{fornengo1997gravitational}
N~Fornengo, C~Giunti, CW~Kim, and J~Song.
\newblock Gravitational effects on the neutrino oscillation.
\newblock {\em Physical Review D}, 56(4):1895, 1997.

\bibitem{kunst2016comparing}
Daniela Kunst, Tom{\'a}{\v{s}} Ledvinka, Georgios Lukes-Gerakopoulos, and
  Jonathan Seyrich.
\newblock Comparing hamiltonians of a spinning test particle for different
  tetrad fields.
\newblock {\em Physical Review D}, 93(4):044004, 2016.

\bibitem{barausse2009hamiltonian}
Enrico Barausse, Etienne Racine, and Alessandra Buonanno.
\newblock Hamiltonian of a spinning test particle in curved spacetime.
\newblock {\em Physical Review D}, 80(10):104025, 2009.

\bibitem{bahcall200610}
John~N Bahcall, Aldo~M Serenelli, and Sarbani Basu.
\newblock 10,000 standard solar models: a monte carlo simulation.
\newblock {\em The Astrophysical Journal Supplement Series}, 165(1):400, 2006.

\bibitem{Kersten:2013fba}
J\"orn Kersten.
\newblock {Coherence of Supernova Neutrinos}.
\newblock {\em Nucl. Phys. B Proc. Suppl.}, 237-238:342--344, 2013.

\bibitem{kersten2016decoherence}
J{\"o}rn Kersten and Alexei~Yu Smirnov.
\newblock Decoherence and oscillations of supernova neutrinos.
\newblock {\em The European Physical Journal C}, 76(6):1--20, 2016.

\bibitem{porto2021coherence}
Yago~P Porto-Silva and Alexei~Yu Smirnov.
\newblock Coherence of oscillations in matter and supernova neutrinos.
\newblock {\em Journal of Cosmology and Astroparticle Physics}, 2021(06):029,
  2021.

\bibitem{akhmedov2017collective}
Evgeny Akhmedov, Joachim Kopp, and Manfred Lindner.
\newblock Collective neutrino oscillations and neutrino wave packets.
\newblock {\em Journal of Cosmology and Astroparticle Physics}, 2017(09):017,
  2017.

\bibitem{wolfenstein1978matter}
L.~Wolfenstein.
\newblock Neutrino oscillations in matter.
\newblock {\em Phys. Rev. D}, 17:2369--2374, May 1978.

\bibitem{mikheev1985resonance}
SP~Mikheev and A~Yu Smirnov.
\newblock Resonance amplification of oscillations in matter and spectroscopy of
  solar neutrinos.
\newblock {\em Yadernaya Fizika}, 42(6):1441--1448, 1985.

\bibitem{fantini2018formalism}
Guido Fantini, Andrea~Gallo Rosso, Francesco Vissani, and Vanessa Zema.
\newblock The formalism of neutrino oscillations: an introduction.
\newblock {\em arXiv preprint arXiv:1802.05781}, 2018.

\bibitem{bjerrum2003quantum}
Niels Emil~Jannik Bjerrum-Bohr, John~F Donoghue, and Barry~R Holstein.
\newblock Quantum corrections to the schwarzschild and kerr metrics.
\newblock {\em Physical Review D}, 68(8):084005, 2003.

\bibitem{calmet2019quantum}
Xavier Calmet, Roberto Casadio, and Folkert Kuipers.
\newblock Quantum gravitational corrections to a star metric and the black hole
  limit.
\newblock {\em Physical Review D}, 100(8):086010, 2019.

\end{thebibliography}


\end{document}